\documentclass[aps,prev,onecolumn,preprintnumbers,floatfix,nofootinbib]{revtex4-1}
\usepackage{graphicx}
\usepackage{bm}
\usepackage{times}
\usepackage{slashed}
\usepackage{color}
\usepackage{eqnarray,amsmath}
\usepackage[T1]{fontenc}
\usepackage{libertine}
\pdfoutput=1

\begin{document}

\title{Observational constraints on the non-flat $\Lambda CDM$ model and a null test using the transition redshift}

\author{Velasquez-Toribio, A.M.}
\email{alan.toribio@ufes.br}
 \affiliation{Departamento de F\'{\i}sica, Nucleo Cosmo-ufes, Universidade Federal do Espirito Santo, 29075-910 Vitøria - ES, Brasil}

\author{Magnago, A. dos R}
\email{armagnago@gmail.com }
\affiliation{Departamento de F\'{\i}sica, Universidade Federal do Espirito Santo, 29075-910, Brasil.}

\begin{abstract}
A natural extension of the standard cosmological model are models that include curvature as a free parameter. In this work we study in detail the observational constraints on the non-flat $\Lambda CDM$ model using the two main geometric tests: SNIa and Hubble parameter measurements. In general we show that the observational constraints on the parameters of the $\Lambda CDM$ model strongly depend on the curvature parameter. In particular, we study the constraints on the transition redshift ($z_{t}$) of a universe dominated by matter for a universe dominated by the cosmological constant. Using this observable we construct a new null test defining $\zeta = z_{t, flat}-z_{t, non flat}$. 
This test depends only on the data of the Hubble parameter, the Hubble constant and the matter density parameter. However, it does not depend on derivative of an observable as generally many tests in the literature. 
To reconstruct this test, we use the Gaussian process method. When we use the best-fit parameters values of $PLANCK/2018$, we find no evidence of a disagreement between the data and the standard model (flat $\Lambda CDM$), but if we use the value $H_{0}$ from $RIESS/2018$ we found a disagreement with respect at the standard model. However, it is important to note that the Hubble parameter data has large errors for a solid statistical analysis.

\end{abstract}

\maketitle

\section{Introduction.}

The standard cosmology model is the $\Lambda CDM$ model with flat spatial curvature. From the theoretical point of view, this model is the simplest explanation of the accelerating expansion of the universe \cite{perlmutter,riess} and is the model that best fits the different types of observational data: SNIa, BAO, CMB, Hubble parameter measurements, etc \cite{carroll, peebles, tanabashi, planck, des}.

On the other hand, recently in the literature some authors have claimed that the standard model does not fit the data
of CMB lensing. Planck's collaboration \cite{planck} found more lensing effect than expected and to quantify this result, the $A_ {lens}$ parameter was introduced. In the reference \cite{silk} it is considered that if the curvature is included, CMB lensing data indicates that the best fit corresponds to a closed universe at more than $99\%$ C.L.

In addition, in the reference \cite{handley}, the author claims a new cosmological tension, curvature tension. The author explicitly shows that the predictions for the curvature parameter using BAO, CMB lensing measurements and data of the $SH0ES$ Collaboration \cite{shoes} are incompatible with each other.
However, in the reference \cite {gratton} the authors using a subsample of data of CMB lensing have confirmed a trend for a closed universe, but, when adding BAO data, the standard model remains the best fit.
Another two inconsistencies with respect at standard model are
the tension of the Hubble constant using CMB data and low redshift data \cite{riesshubble} and the data of cosmic shear of $KiDs-450$ measurements \cite{shear}.

Independently if future observations demonstrate that the data are compatible with the flat $\Lambda CDM$ model this current polemic places to the curvature parameter in an important position in the cosmological discussion.
From the historical point of view the question of introducing the curvature parameter in a cosmological model is old, can be traced back to Einstein and De Sitter \cite{einstein}. In this article they considered a Universe with finite matter density, in a homogeneous and isotropic model.
The authors conclude that the observational data available at the time do not imply introducing the curvature, but mention that future data should allow limiting curvature values.

On the other hand, in recent times, in the literature has been studying different aspects of the curvature, such as: the equivalence between 
$\Lambda CDM$ models with curvature and dynamic dark energy models \cite{ichikawa,ichikawai,ichikawaii,ichikawaiii}. The influence of the curvature on the observational 
constraints of the equation of state \cite{indepen}. Also the model-independent approaches has been used to place observational constraints on the curvature \cite{indepenn}, among other considerations.

In addition to the curvature a phenomenological parameter that allows to characterize the accelerated expansion of any cosmological model is the deceleration parameter, which in turn allows us to study the redshift of the transition from a decelerated universe to an accelerated universe. This parameter has been quite studied in the flat $\Lambda CDM$ model\cite{waga}. However, the determination of the transition redshift including curvature as a free parameter has been little investigated, with the exception of references \cite{ratra,ratraa}.

Motivated by all these recent results, in the present work, we study observational constraints on the transition redshift in the
non-flat $\Lambda CDM $ model using two geometric tests: SNIa data and Hubble parameter measurements.

Considering that the transition redshift in the non-flat $\Lambda CDM$ model has an analytical expression, we can rewrite the Hubble parameter directly as a function of the transition redshift and the curvature parameter. In this way we avoid an extra 
propagation of errors. Additionally, it is important to note that we get quite general results, 
because to construct the confidence contours, we do not fix the values of the parameters, but use a marginalization process in a given range of values for the parameters.

Also, using the concept of transition redshift, we propose a new null test, which for the non-flat $\Lambda CDM$ model
explicitly depends on the Hubble parameter, the Hubble constant and the matter density parameter today. We use observational data from the Hubble parameter to reconstruct the expression of this test. As a statistical method we use the non-parametric method of Gaussian processes.
We demonstrate that this test is strongly sensitive to the values of the parameters cosmological: $(\Omega_ {m0}, H_ {0})$.

Our paper is organized as follows. In Sec. 2 we summarize the cosmological dynamics of the non-flat $\Lambda CDM$ model. In Sec 3. we present a new null test using the transition redshift in Sec. 4 we present our data and in the Sec. 5 we present our result and conclusions.

\section{The $\Lambda CDM$ model in the Background}
Considering the cosmological principle, the FLRW metric can be written as\cite{flrw,flrwl,flrwr,flrww},
\begin{equation}
ds^{2} = -dt^{2}+ a(t)^{2}\left[\frac{dr^{2}}{1-kr^{2}}+r^{2}d\theta^{2} +r^{2} \sin^{2} \theta d\phi^{2}\right],
\end{equation}
where $a(t)$ is the scale factor and $k$ is the spatial curvature which can be $k = + 1$ for a closed universe, $k=0$ for a flat Universe and $k = -1$ for an open universe. In addition, if we consider Einstein's equations and a tensor energy-momentum of perfect fluid, then we can derive the fundamental equation of cosmology \cite{weinberg},

\begin{eqnarray}
H^{2} = (\frac{\dot{a}}{a})^{2} = \frac{8 \pi G \rho}{3} + \frac{\Lambda}{3} - \frac{k}{a^{2}}.
\end{eqnarray}

This equation can be rewritten using the redshift as:
\begin{eqnarray}
H = H_{0}\sqrt{\Omega_{m0}(1+z)^{3}+ \Omega_{k0}(1+z)^{2} + \Omega_{\Lambda 0}},
\end{eqnarray}
where we use the definitions:

\begin{equation}
\begin{array}{lll}
\Omega_{m0} = \frac{8\pi G \rho_{m0}}{3H_{0}^{2}}, & \Omega_{\Lambda 0} = \frac{\Lambda}{3H_{0}^{2}} &  {\rm and}~\Omega_{k0} = \frac{-k}{a^{2}H^{2}}.
\end{array}
\end{equation}

\begin{figure*}[htbp] 
 	\centering
	\includegraphics[scale=0.700]{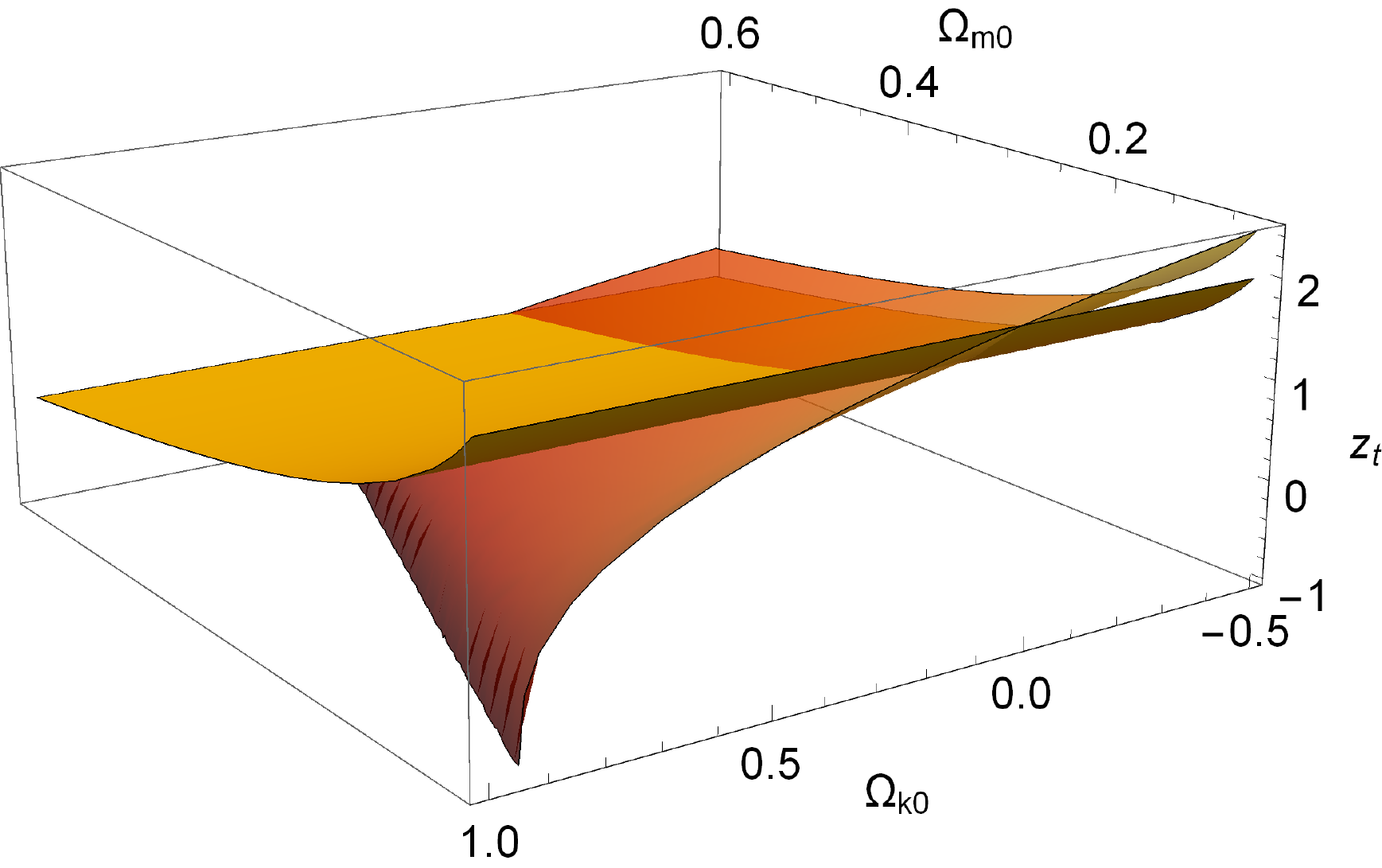}
 	\caption{The curved surface represents the transition redshift as a function of the material and curvature parameters in the non-flat $\Lambda CDM$ model and the other curve represents the transition redshift of the flat $\Lambda CDM$ model.}
 	\label{fig01}
 \end{figure*}

Additionally, we have the restriction:

\begin{eqnarray}
\Omega_{m0}+\Omega_{\Lambda 0}+\Omega_{k0} =1.
\end{eqnarray}

To calculate the transition redshift, $z_{t}$, of a decelerated to accelerated universe we use the definition of the deceleration parameter,
\begin{eqnarray}
q(z) = -\frac{\ddot{a}}{aH^{2}} = \frac{d}{dt}(\frac{1}{H})-1.
\end{eqnarray}

Thus, using the definition of $H(z)$ and the condition for the transition redshift $q(z_{t})=0$. We can determine that \cite{velasquezt}, 
\begin{eqnarray}
z_{t} &=& \left( \frac{2\Omega_{\Lambda0}}{\Omega_{m0}} \right)^{1/3}  -  1  \\
      &=& \left(\frac{2 \left(1-\Omega_{m0}-\Omega_{k0} \right)}{\Omega_{m0}}\right)^{1/3}  -  1,
\end{eqnarray}
where we can observe that the transition redshift for a non-flat universe 
$\Lambda CDM$ is a analytical function of the parameters of relative densities \cite{velasquezt}.
Using equations (5) and (7) we can explicitly rewrite the hubble parameter using the variables, 
($\Omega_{k0}$ e $z_{t} $)as,

\begin{equation}
H = H_{0} \sqrt{\frac{(1-\Omega_{k0})(1+z)^{3}}{\frac{1}{2}(1+z_{t})+ 1} + \Omega_{k0}(1+z)^{2} +\frac{(1-\Omega_{k0})(1+z_{t})^{3}}{2(\frac{1}{2}(1+z_{t})+ 1)}}.
\end{equation}
We can also use the variables $(\Omega_{m0}, z_{t} )$ to write:

\begin{eqnarray}
H = H_{0} \sqrt{\Omega_{m0}(1+z)^{3} + 
(1-\Omega_{m0}(1+\frac{(1+z_{t})^{3}}{2}))(1+z)^{2} + \frac{\Omega_{m0}}{2}(1+z_{t})^{3}}.
\label{eqztom}
\end{eqnarray}

This expression of the Hubble parameter is important because the statistics of the $\chi^{2}$ constructed will explicitly depend on these parameters avoiding a extra propagation of errors.
The importance of the curvature parameter in determining the transition redshift can best be observed by means of Figure 1. where we show the function $z_ {t}(\Omega_ {m}, \Omega_ {k0})$. This function is a well-behaved three-dimensional surface, except for extreme value of the curvature and very low values for the matter. But these regions are excluded by observational data. In the figure the intersection of the planes corresponds to the case where the curvature is zero. A quick inspection allows us to observe that if we consider curvature, then there are different ways to accommodate the measures on the surface of $z_ {t}$. In particular if we consider that the observations various determine $z_ {t}$ preferably in the range $ (0.5-1.00) $, then the inclusion of the curvature allows that the value of $z_ {t}$ can easily be accommodated outside of the line for a flat universe. In this work we study the constraints associated with this degeneracy employing SNIa and Hubble parameter measurements.

\section{Null test for $\Lambda CDM$}
We can define a new null test to distinguish between flat and non-flat $\Lambda CDM$ models using the redshift transition
To do this we put in evidence the $z_ {t}$ of the expression for the Hubble parameter given by the equation (10) obtaining,
\begin{eqnarray}
z_{t,non-flat}=\left(\frac{\left(\frac{H}{H_{0}}\right)^{2}-\Omega_{m0}(1+z)^{3}-\left(1-\Omega_{m0}\right)(1+z)^{2}}{\Omega_{m0}\left(1-(1+z)^{2}\right)}\right)^{1/3} - 1,
\end{eqnarray}
In an analogous way we can determine for flat transition redshift, so the null test can be formulated as:
\begin{eqnarray}
\zeta = z_{t,flat} - z_{t,nonflat},
\end{eqnarray}
where we can see that if $\zeta = 0$, then the flat model is preferred, otherwise the model with curvature is preferred.
In an explicit way we can write this null test as,

\begin{eqnarray}
\zeta = \left(\frac{(\frac{H}{H_{0}})^{2}-\Omega_{m0}(1+z)^{3}}{\Omega_{m0}}\right)^{1/3}-\left(\frac{(\frac{H}{H_{0}})^{2}-\Omega_{m0}(1+z)^{3}-(1-\Omega_{m0})(1+z)^{2}}{\Omega_{m0}(1-(1+z)^{2})}\right)^{1/3}.
\end{eqnarray}

Interestingly, our test includes the reconstructed data of the Hubble parameter and does not include derivatives
of data such as other tests, for example \cite{estadistica}. Data derivatives in general are difficult to obtain and spread the error remarkably. However, our test explicitly includes the Hubble constant, $H_{0}$, and the matter density parameter today $\Omega_{m0}$.

To reconstruct the observable, $H(z)$, we use only data from the Hubble parameter and as a statistical method we use
the non-parametric method of Gaussian processes. This method is suitable for this case, since it does not assume a specific model to reconstruct the function $H(z)$. Once the $H(z)$ function is obtained, the null test can be reconstructed. As a first approximation to determine the errors of the $\zeta$ function we can use the theory of error propagation,

\begin{eqnarray}
\sigma_{\zeta} &=& \sqrt{\frac{\partial \zeta}{\partial H} \delta H+\frac{\partial \zeta}{\partial H_{0}}\delta H_{0}+\frac{\partial \zeta}{\partial \Omega_{m0}} \delta \Omega_{m0}}
\end{eqnarray}

To perform the reconstruction of Gaussian processes we use the popular public package $GaPP$, which has been applied to a
large number of cosmological studies. For package details you can see the references \cite{gapp}. For a recent use of this package to see the reference \cite{velasquez}.

\begin{table}\center
\begin{tabular}{|c|c|c|}
\hline
Parameters & Best-fitting & Marginalization range \\ \hline \hline
$\Omega_{m0}$ &  $0.320 \pm 0.0300$   &  $0.07 < \Omega_{m0} < 0.600$  \\ \hline
$H_{0}$        &  $69.05 \pm 2.00$     &           $50 < H_{0} < 80$    \\ \hline
$z_{t}$        &  $0.700 \pm 0.300$    &      $0.400 < z_{t} < 1.00$     \\ \hline
$\Omega_{k0}$  &  $0.03 \pm 0.050$     &   $-0.20 < \Omega_{k0} < 0.20$ \\ \hline
\hline
\end{tabular}
\caption{Best-fitting parameters $1\sigma$ confidence intervals.}
\end{table}

\section{Observational Constraints}
For determine observational constraints on the non-flat $\Lambda CDM$ model it is essential to define the comoving distance as,
\begin{eqnarray}
r(z) = \frac{c}{H_{0}}\int_{0}^{z}{\frac{dz'}{E(z')}}.
\end{eqnarray}
To determine the luminosity distance including curvature is necessary to distinguish three cases, for this we define
the transversal comoving distance, $r_{t}$, \footnote{Here we follow the notation of the article by D. Hogg see reference astro-ph/9905116v4}

\begin{eqnarray}
r_{\rm t} = \left\{
\begin{array}{ll}
\frac{c}{H_{0}}\,\frac{1}{\sqrt{\Omega_{k0}}}\,\sinh\left[\sqrt{\Omega_{k0}}\,\frac{H_{0}}{c}r(z)\right] & {\rm for}~\Omega_{k0}>0 \\
r(z) & {\rm for}~\Omega_{k0}=0 \\
\frac{c}{H_{0}}\,\frac{1}{\sqrt{|\Omega_{k0}|}}\,\sin\left[\sqrt{|\Omega_{k0}|}\,\frac{H_{0}}{c}r(z)\right] & {\rm for}~\Omega_{k0}<0
\end{array}
\right.
\end{eqnarray}
using the above definitions we can determine the luminosity distance as,
\begin{eqnarray}
d_{L} = (1+z)r_{t}
\end{eqnarray}

\subsection{Supernovae Ia}
In this study we use the data from Supernovas Ia called "Pantheon" sample \cite{scolnic} which is the largest
combined sample of SNIa and consists of $1048$ data with redshifts in the range $0.01 < z < 2.3$.
 It is a collection of SNe Ia discovered by the Pan-STARRS1 (PS1) Medium Deep Survey
and SNe Ia from Low-z, SDSS, SNLS and HST surveys.
This supernova Ia compilation uses The SALT 2 program to transform light curves into distances using a modified version of the Tripp 
formula \cite {tripp},
\begin{equation}
\mu = m_{B} - M + \alpha x_{1}-\beta c + \Delta_{M} + \Delta_{B},
\end{equation}
where $\mu$ is the distance modulus, $\Delta_{M}$ is a distance correction based on the host-galaxy mass of the SNIa and $\Delta_{B}$
is the distance correction based on predicted bias from simulations. Also $\alpha$ is the coefficient of the relation between luminosity and stretch, $\beta$ is the coefficient of the relation between luminosity and color and $M$ is
the absolute $B$-band magnitude of a fiducial SNIa with $x_{1} = 0$ and $c = 0$. Also $c$ is the color and $x_{1}$ is the light-curve shape parameter and $m_{B}$ is the log of the overall flux normalization.
An uncertainty matrix $\bf{C}$ is defined such that,
\begin{equation}
\chi_{SNIa}^{2} = \Delta \vec{\mu}^{T}. \mathbf{C}^{-1}. \Delta \vec{\mu},
\end{equation}
where $\Delta \vec{\mu} = \vec{\mu}_{obs}- \vec{\mu}_{model}$ and $\vec{\mu}_{model}$ is a vector of distance modulus from a given cosmological model and $\vec{\mu}_{obs}$ is a vector of observational distance modulus. The $\vec{\mu}=\vec{m}-M$, where $M$ is the absolute magnitude and $\vec{m}$ is the apparent magnitude, which is is given by
\begin{equation}
\vec{m}_{model} = M+5Log_{10}(D_{L}) + 5Log_{10}(\frac{c/H_{0}}{1Mpc}) + 25 = \bar{M}+25+5Log(D_{L}).
\end{equation}
where $D_{L}=\frac{H_{0}}{c} d_{L}$ and $\bar{M} = M+5 \log_10 (\frac{c/H_{0}}{1Mpc})$ is an nuisance parameter, which depends on the Hubble constant $H_{0}$ and the absolute magnitude $M$. To minimize with respect to the nuisance parameter we follow a process similar at the references \cite{scolnic,conley}. Therefore the $\chi^{2}_{\bar{M},marg}$ is,
\begin{equation}
\chi^{2}_{\bar{M}, marg} = a+ \log_10 ( \frac{e}{2\pi} ) -\frac{b^{2}}{e},
\end{equation}
where,
\begin{eqnarray*}
a&=&\Delta \vec{m}^{T}.C^{-1}.\Delta \vec{m},\\
b&=&\Delta \vec{m}^{T}.C^{-1}.I,\\
e&=&I^{T}.C^{-1}.I
\end{eqnarray*}
where $\Delta \vec{m}= \vec{m}_{obs}-\vec{m}_{model}$ and $I$ is the identity matrix.

\subsection{Hubble Parameter Measurements}
There are two efficient and widely used forms to obtain Hubble parameters measurements:
\begin{itemize}
\item Cosmic Chronometers(CC): This method is based on the expression of the differential age of the
universe as a function of redshift,
\begin{equation}
 H(z) = -\frac{1}{1+z}\frac{dz}{dt}.
\end{equation}
This method was proposed by directly measuring the amount $dz/dt $ and, consequently, 
the Hubble parameter. The most used data to measure this amount have been passively evolving galaxies 
with high-resolution spectroscopic data along with synthetic catalogs to limit the age 
of the oldest stars in the galaxy. A complete description of this methodology can be reviewed 
for the SDSS in the reference \cite{hubble}.

\item The Radial BAO Size Method:This method is based on measurements of the scale of $BAO$.
This method is more accurate with respect to $CC$. 
This accuracy is understandable because $BAO$ mainly depends on a spatial 
measurement compared to the first method where a time measure is 
required which increases systematic errors. However, this method of $BAO$ 
requires assuming a prior in the radius of the sound horizon,$r(z)$, so that
\begin{equation}
 H(z) = -\frac{r_{bao}(z)}{r_{cmb}(z)}H_{fiducial}(z).
\end{equation}
This method depends on the fiducial model, usually the model associated with $CMB$, is the $\Lambda CDM$ model.
\end{itemize}
In the literature there are different compilations of samples of the Hubble parameters data, we use the sample presented by 
\cite{marek} what includes data of $CC$ and $BAO$. These data cover a range $0.2 <z<2.35$ in redshift.
We can construct the statistics $\chi^{2}_{H}$ as,
\begin{equation}
\chi^{2}_{H}(\Omega_{m0},\Omega_{\Lambda},H_{0}) = \sum_{i=1}^{36}{\frac{(H_{obs,i}-H_{model}(z_{i},\Omega_{m0},\Omega_{\Lambda},H_{0}))^{2}}{\sigma_{obs,i}^{2}}},
\end{equation}
where $H_{obs,i}$ are the observational data and $H_{model}$ are the theoretical values determined by equation (3) and the $\sigma_{obs,i}$ are the errors of the observational data.

\subsection{Combining Data}
We combine the data by adding the $\chi^{2}$ of each dataset, so we get
\begin{equation}
\chi^{2}_{total}(\Omega_{m0},\Omega_{\Lambda 0},H_{0}) = \chi^{2}_{SNIa, marg} + \chi^{2}_{H}.
\end{equation}

We can construct the probability contours through the marginalization process, thus, for example, for the case
of $(\Omega_{m0},\Omega_{\Lambda 0})$ we integrate on the likelihood with respect $H_{0}$,

\begin{equation}
L(\Omega_{m0},\Omega_{\Lambda 0}) = -2 Log_{10} \left[\int_{-80}^{60}{e^{-\frac{\chi_{total}(\Omega_{m0},\Omega_{\Lambda 0},H_{0})^{2}}{2}} dH_{0}}\right].
\end{equation}
For other sets of parameters we proceed analogously.

\section{Results and discuss}

We investigate the observational constraints on the different combinations of cosmological parameters. In Figure 2, we show the observational constraints using SNIa and the Hubble parameter data. In all cases we use a marginalization over the Hubble parameter in the range 
$50 <H_ {0} <80$. The figure on the right side show the constraints due to the sum of the data. It is worth mentioning the lower right figure, which shows the dependence between the curvature parameter and the transition redshift. But it is also evident that the flat model is within $ 1 \sigma $. The figure 3, on the left side, we show the strong degenescescence between the matter density parameter and the parameter
of curvature for SNIa data, in the middle figure we show the confidence contours for the matter versus the transition redshift. This result updates the calculation shown in the reference \cite{lima}. The discontinuous curve represents the flat universe.

In Figure 4 we show the dependence of the parameters with respect to the Hubble constant, $ H_ {0}$. For all cases we use a marginalization interval on the matter density parameter of $ 0.07 <\Omega_ {m0} <0.6 $. On the right side we show the constraints of the sum 
$ SNIa + H$. We can see that the variation range of, $z_{t}$, corresponds to $ 0.4 <z_ {t} <1.00 $ and of the curvature parameter for $ -0.2 <\Omega_ {k0} <0.2 $. In Figure 5, we show the PDFs for all the parameters studied.
The best-fitting values are shown in Table I with $ 1 \sigma $. In Figure 6, we show the three-dimensional constraints for
the parameter space $(\Omega_ {k0}, z_{t}, H_{0}) $. In this figure we shown that the geometric tests can restrict the parameters values on a closed surface for the non-flat $\Lambda CDM $ model, but the flat model $\Lambda CDM$ cannot be ruled out.

In figure 7. we shown the result of the null test using the Gaussian process method. In the figure on the left side we use the best-fitting values of the $PLANCK$/2018 \cite{planck}, $ H_ {0} = 67.4 \pm 0.5$
and $ \Omega_ {m0} = 0.315 \pm 0.007$. We can see that the flat $\Lambda CDM$ model adjusts the null test very well.
However, if we use the local value of the Hubble constant of $RIESS$ et al./ 2018\cite{riesshubble} $H_{0} = 74.03 \pm 1.42$ and keep the same value of $\Omega_ {m0} $ the null test at $2\sigma$ does not correspond to the flat $\Lambda CDM $ model. On the other hand, if we use a smaller value of the parameter of matter, maintaining the value of $ H_ {0} $ of $RIESS$ et al., we can reconcile the flat $\Lambda CDM$ model in $1\sigma$.

\section{Conclusion}

In the present work we have investigated the observational constraints on the non-flat $\Lambda CDM$ model using as observational data the SNIa and Hubble data. We have emphasized the study on the transition redshift, which can be used to construct a null test. This test is sensitive to the values of $H(z) $, $ H_{0} $ and $\Omega_{m0} $, but does not include derivatives of cosmological observables, which prevents excessive propagation of errors.

In general, we have shown a strong dependence between the constraints of the curvature parameter and all other parameters.
Our results are quite general, because we do not use fixed values of the parameters, but we have used a marginalization process integrating in a large interval to allow considerable changes of the parameters. In all case the flat $\Lambda CDM$ model cannot be excluded. A subsequent study may include other observables such as $BAO$, $QSO$ and structure growth data to constraints
the transition redshift with curvature, i.e., the $ (z_{t},\Omega_{k0}) $ plane.

On the other hand, the null test is quite sensitive to the values of the parameters, for the best-fitting of Planck/2018 the flat $\Lambda CDM$model is preferred, but if use the Hubble constant value local of $Riess$ et al. 2018 our null test excludes the flat model with 
$2\sigma$. If we consider values less for the matter content, for example, $ \Omega_{m0} = 0.28$, we can alleviate the rejection of the flat model at $ 1\sigma$. To reconstruct the null test we only use data from the Hubble parameter measurements and the non-parametric method of Gaussian processes. We do not include $SNIa$, since it would imply reconstruct derivatives and would spread the error excessively.
Our null test may be interesting to study other cosmological models, such as a quintessence model or interaction between 
dark energy and dark matter models.

\begin{figure*}[htbp] 
 	\centering
	\includegraphics[scale=0.300]{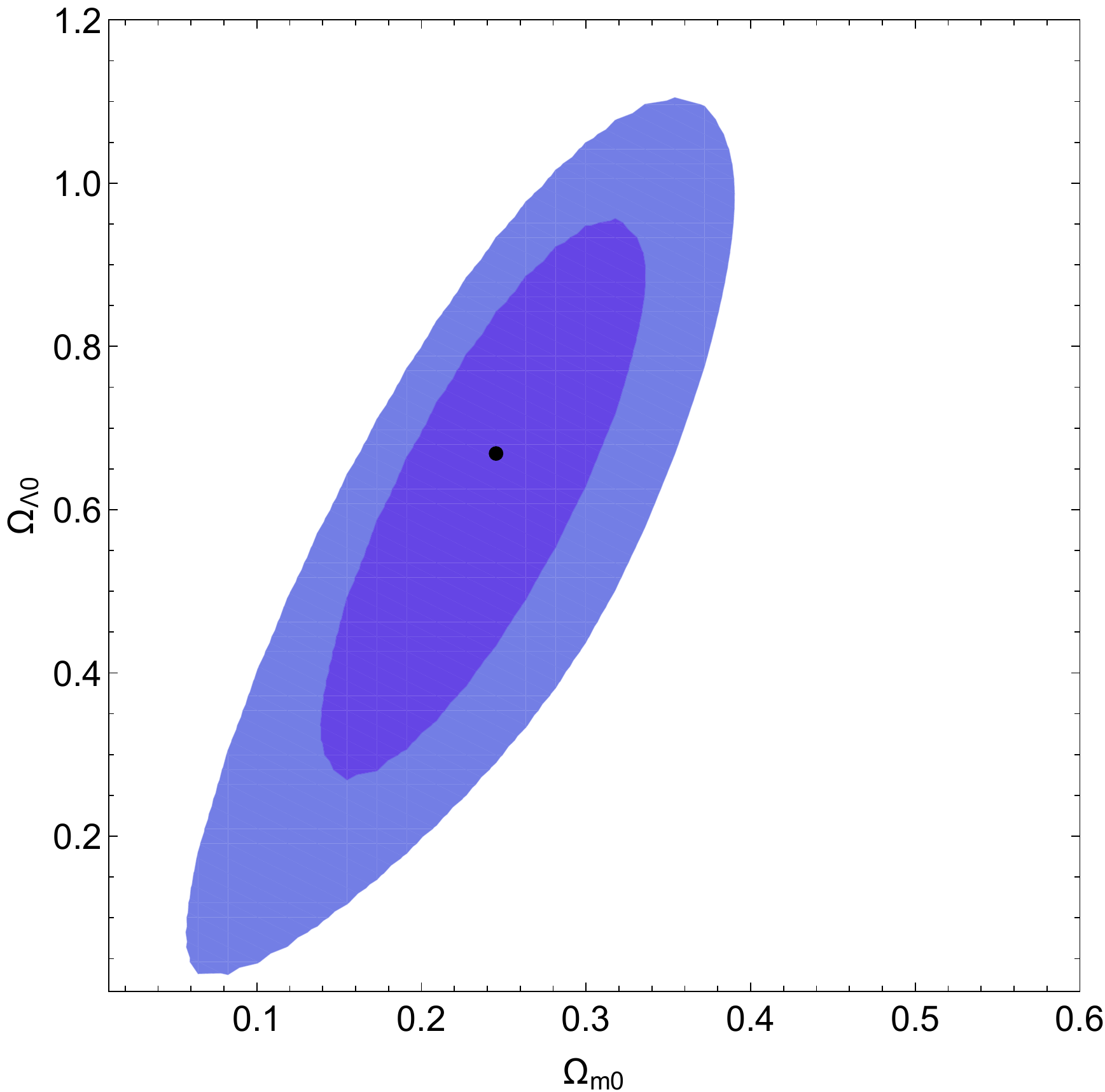}
	\includegraphics[scale=0.300]{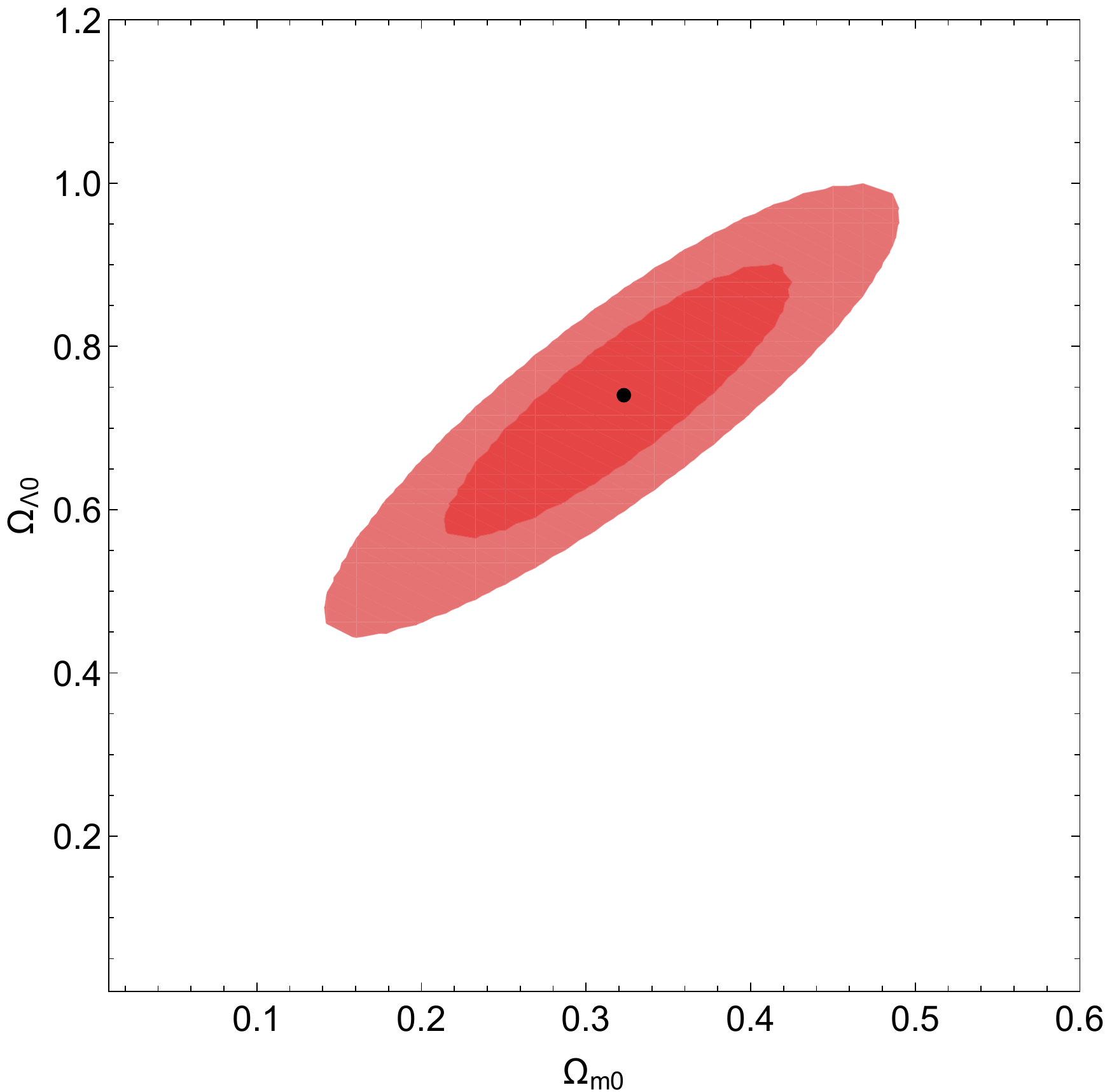}
 	\includegraphics[scale=0.300]{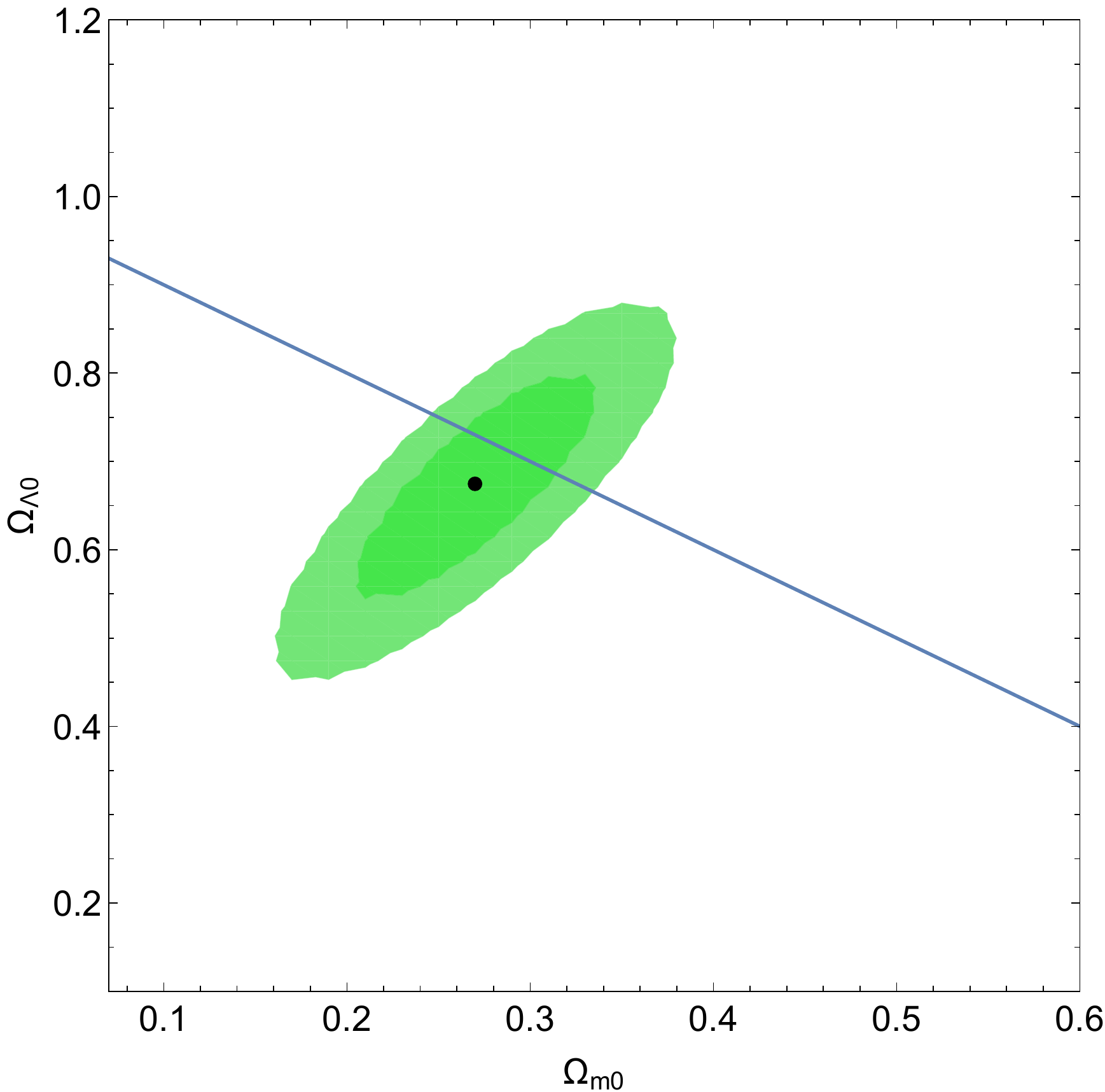}
	\includegraphics[scale=0.300]{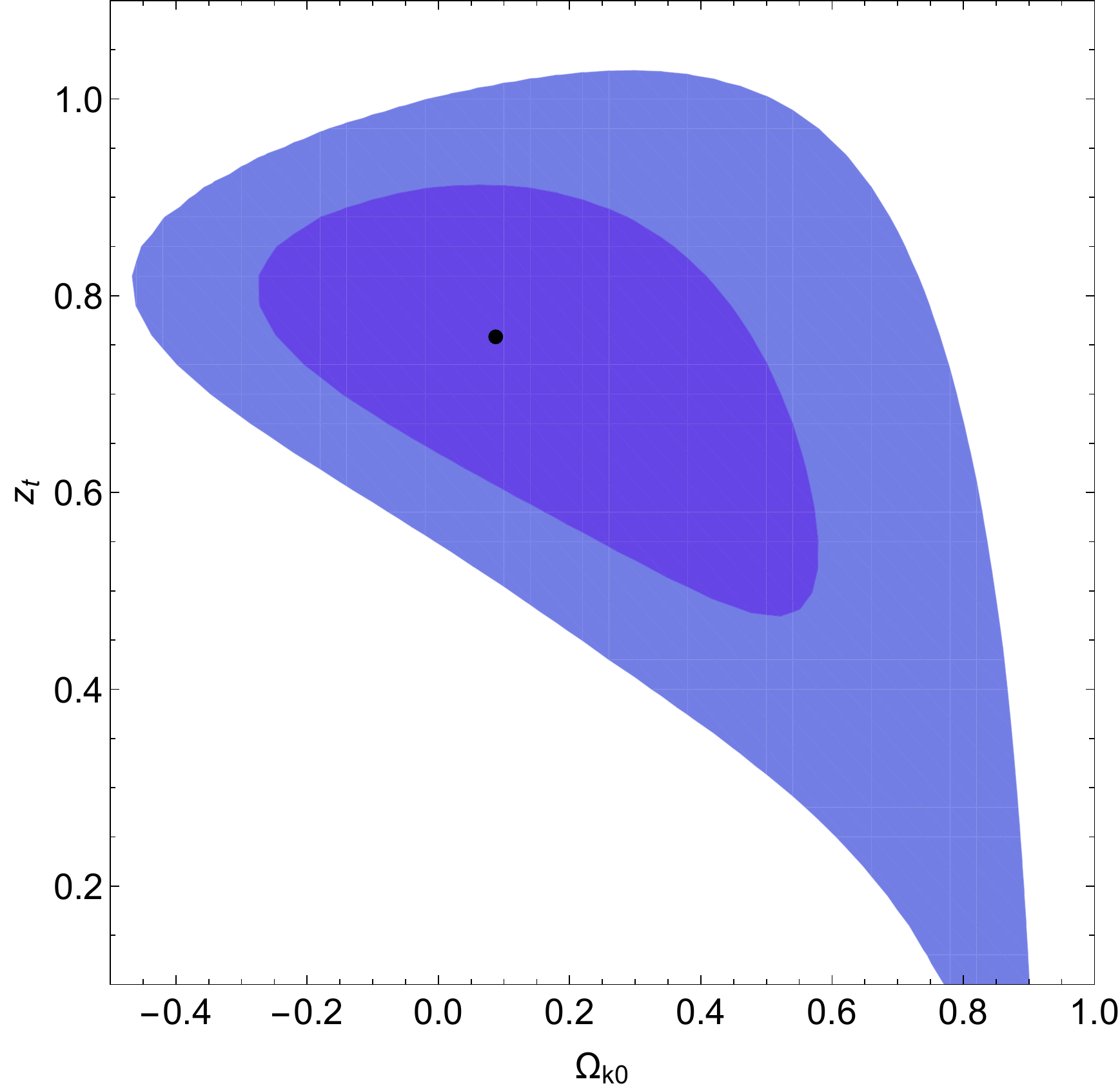}
	\includegraphics[scale=0.300]{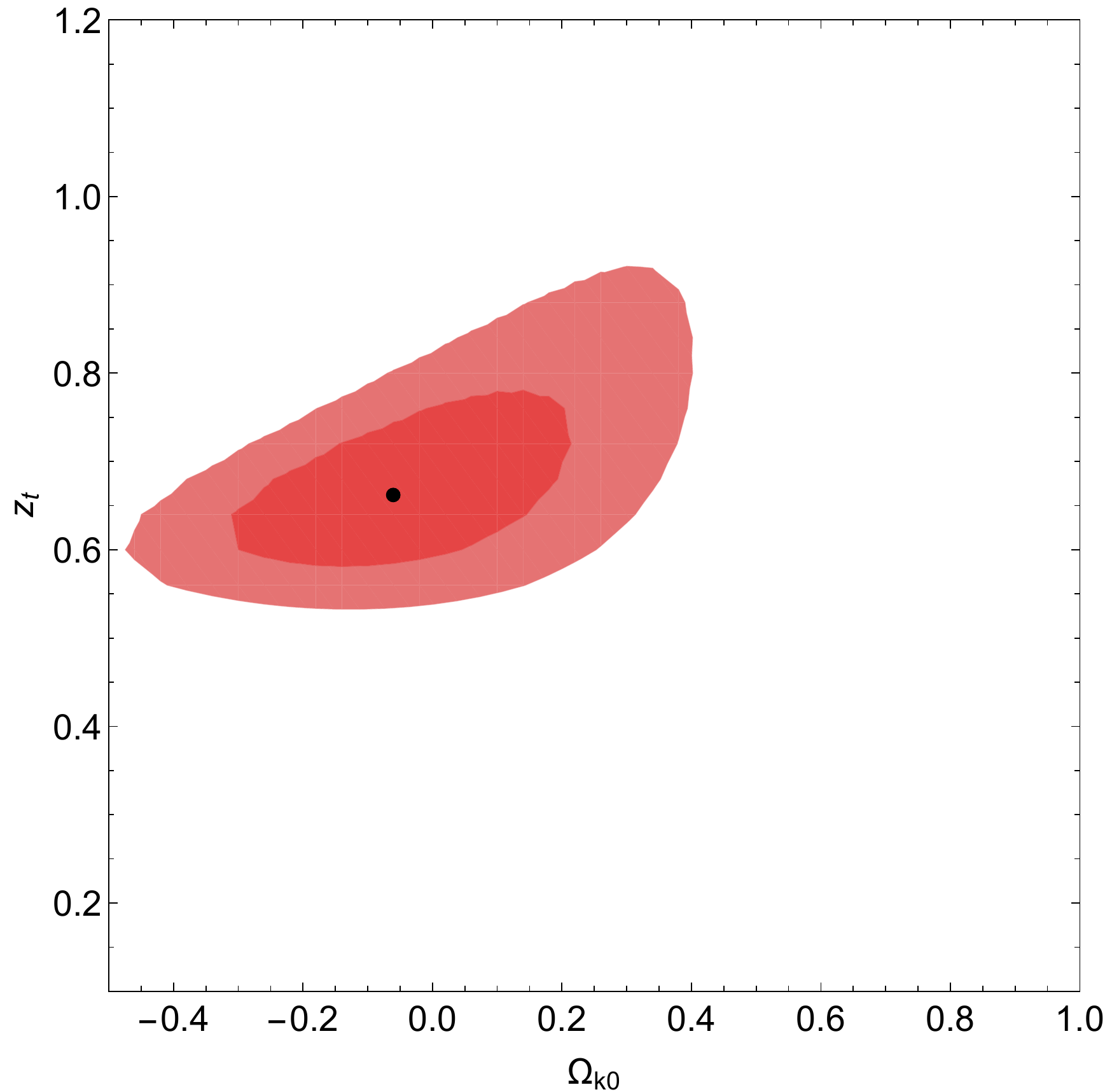}
 	\includegraphics[scale=0.300]{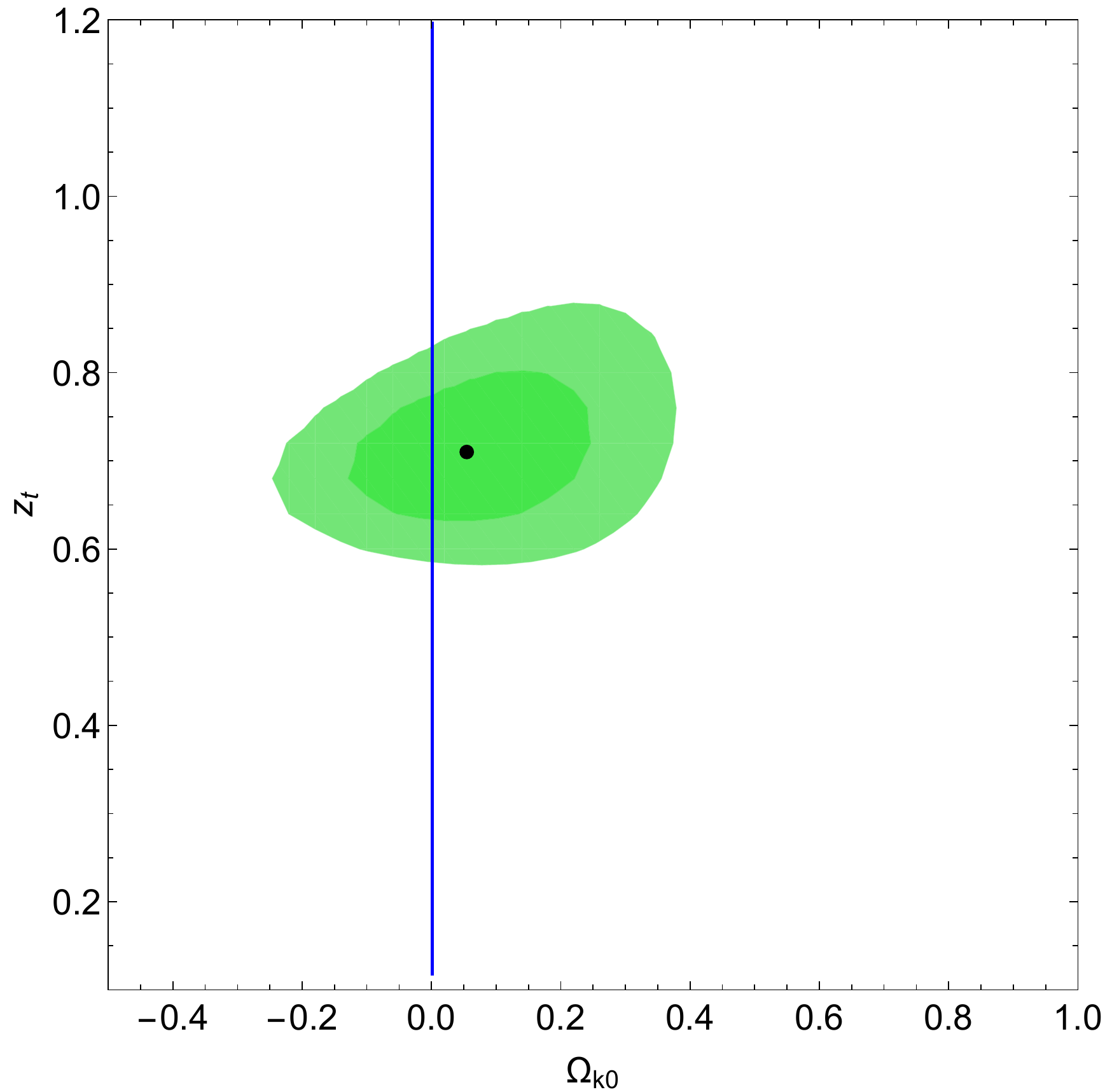}
	
 	\caption{Observational constraint using Hubble parameters measurements (left) and SNIa(middle). A joint analysis is shown on the right side. In all case we marginalized $50<H_{0}<80) $. The blue line represents the universe with flat curvature.}
 	\label{fig01}
 \end{figure*}

\begin{figure*}[htbp] 
 	\centering
	\includegraphics[scale=0.300]{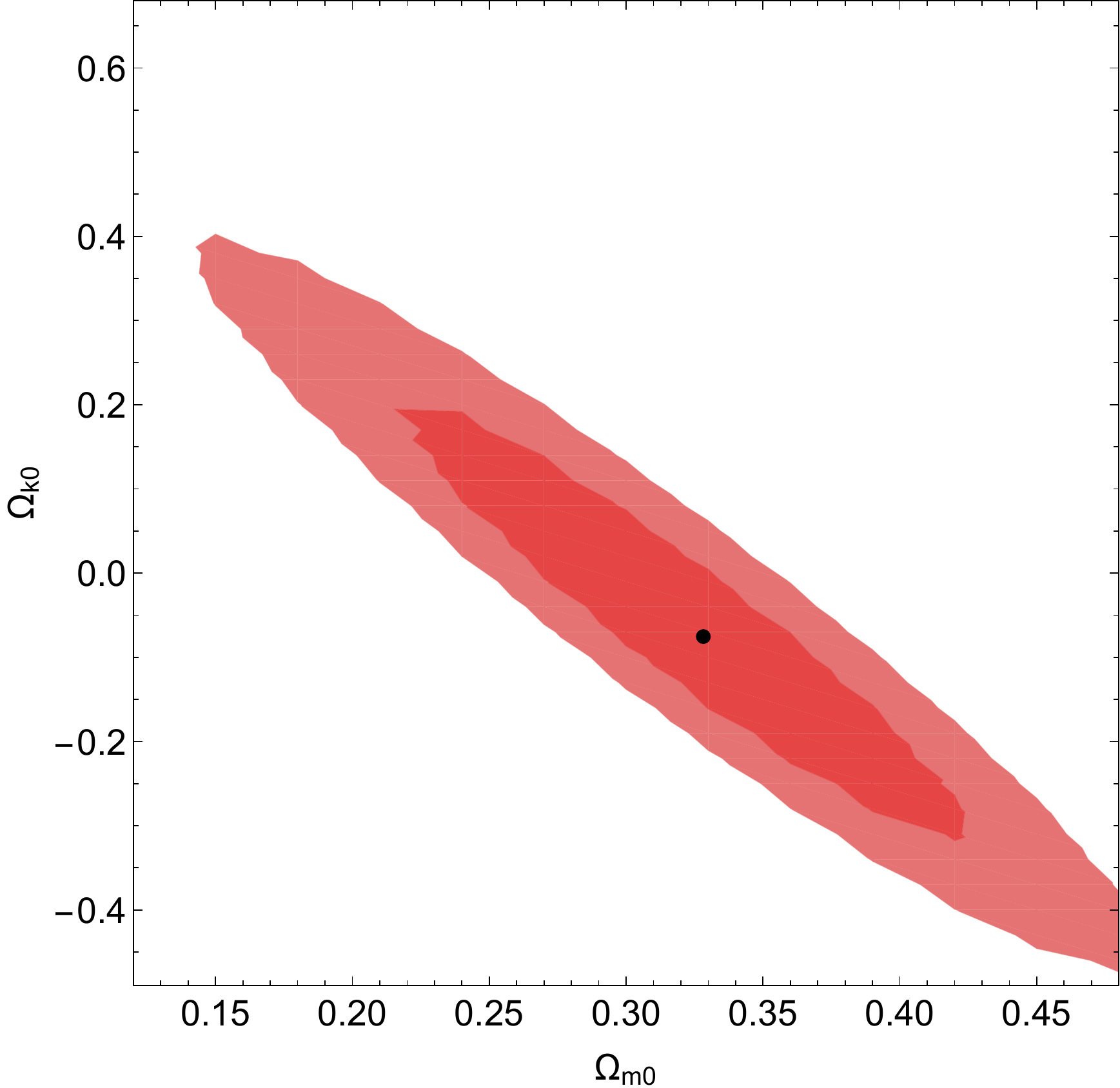}
	\includegraphics[scale=0.300]{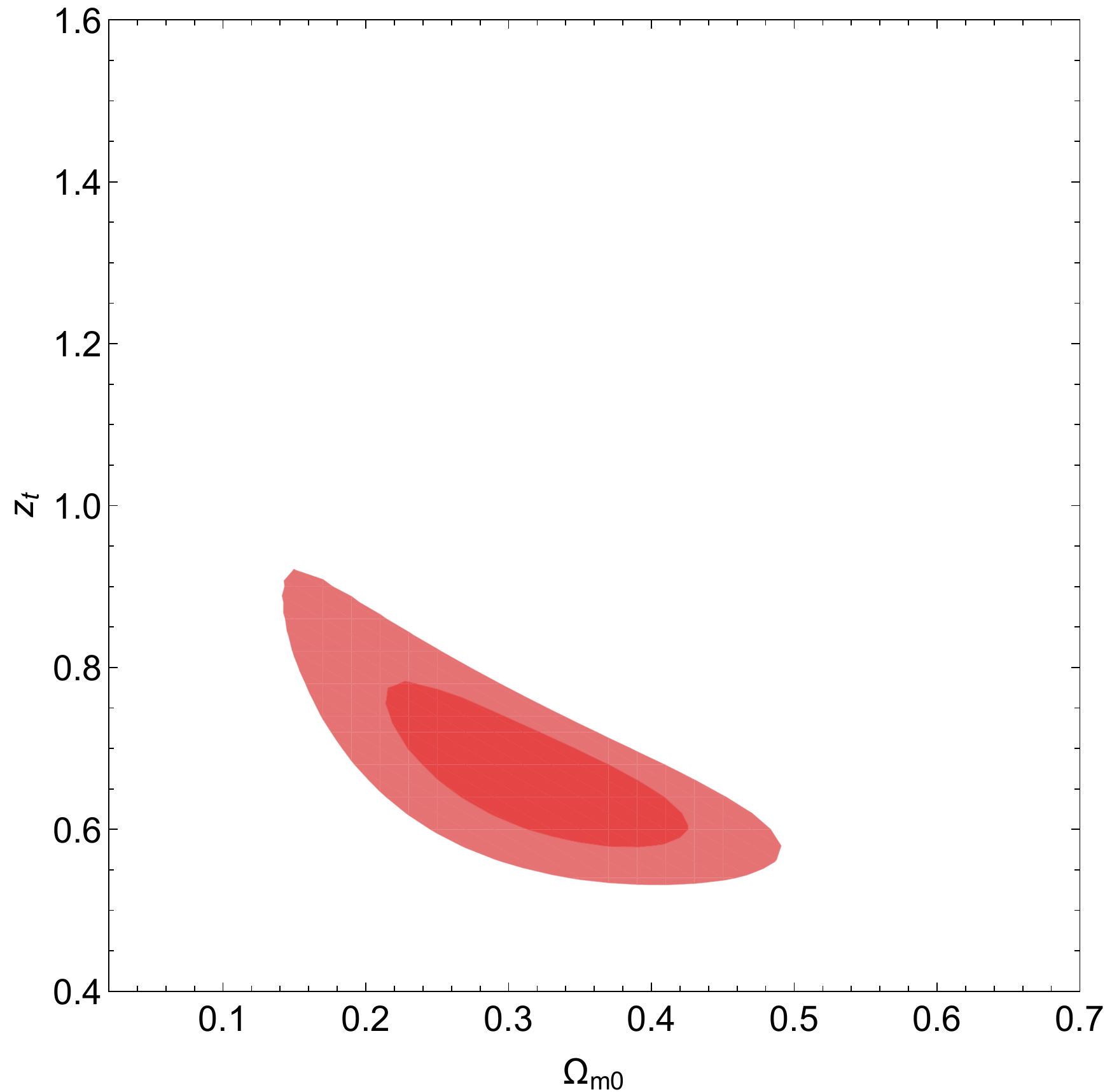}
	\includegraphics[scale=0.300]{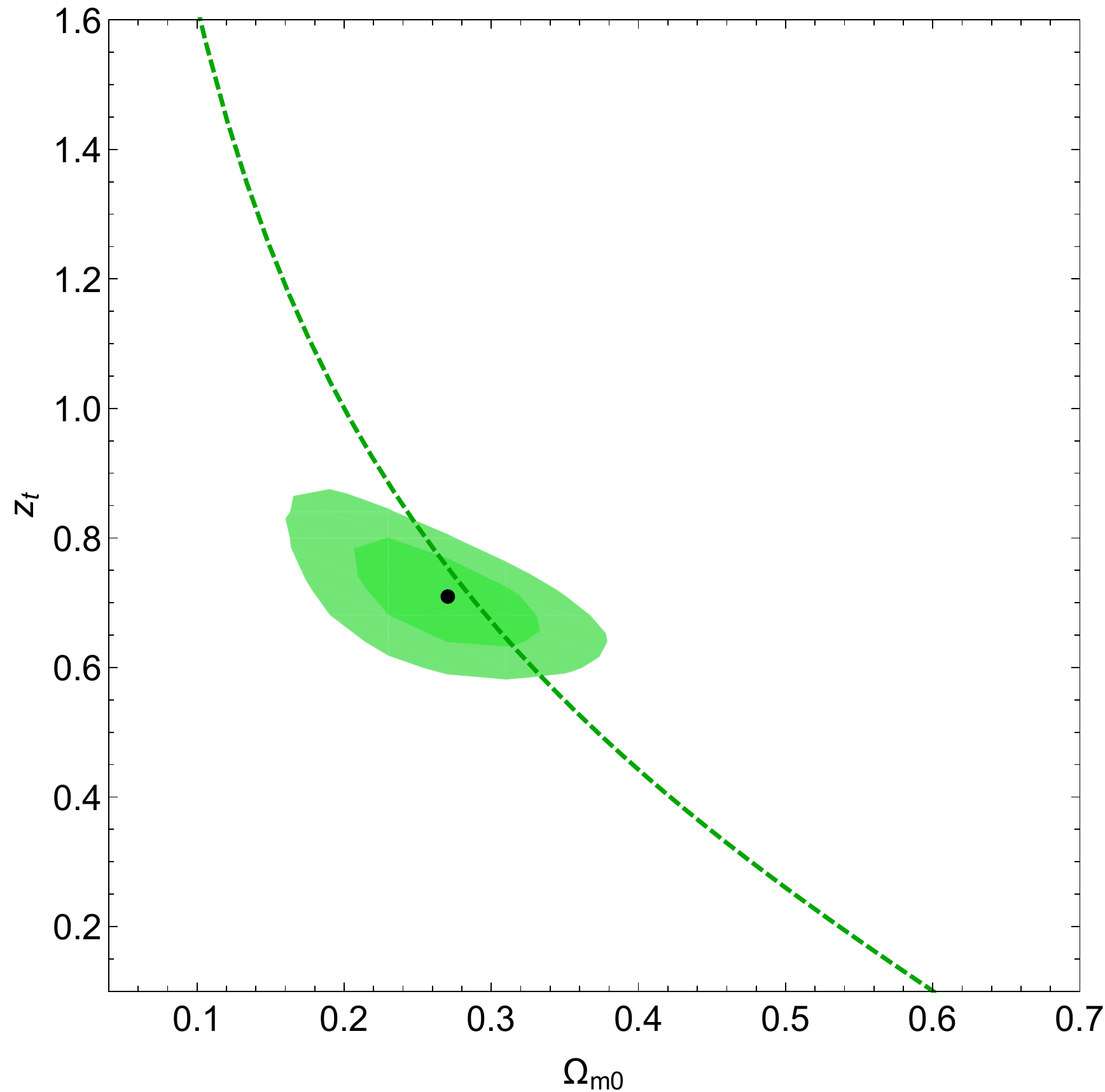}
 	\caption{In the figure on the left we show a strong degenerescence between the curvature parameter and the matter density parameter. In the central figure we show observational constraints updated for the parameter space studied in the reference \cite{lima} and on the right we show the joint data observational constraints. The green dashed line represents the universe with flat curvature.}
 	\label{fig01}
 \end{figure*}

\begin{figure*}[htbp] 
 	\centering
	\includegraphics[scale=0.400]{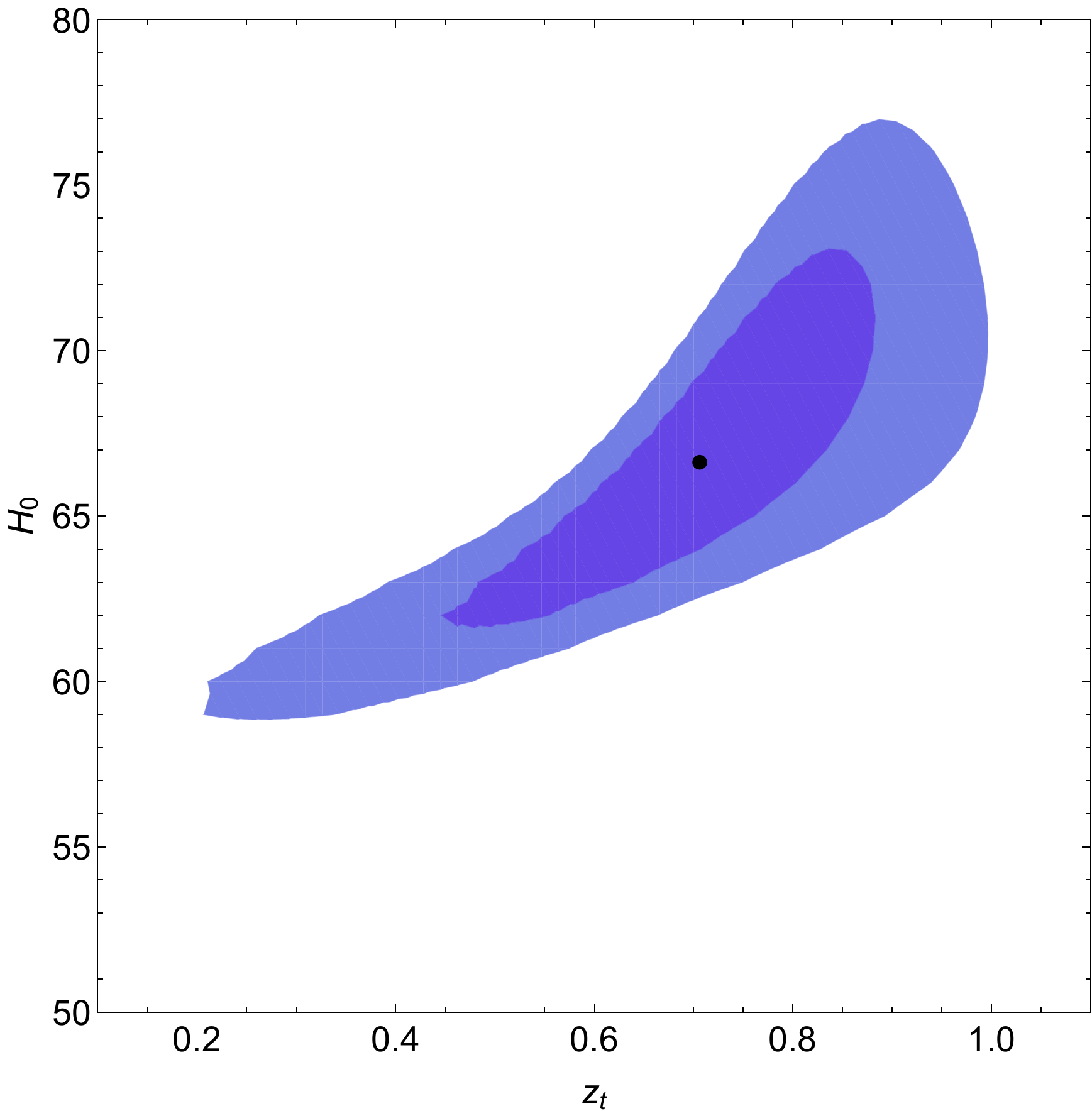}
 		\includegraphics[scale=0.400]{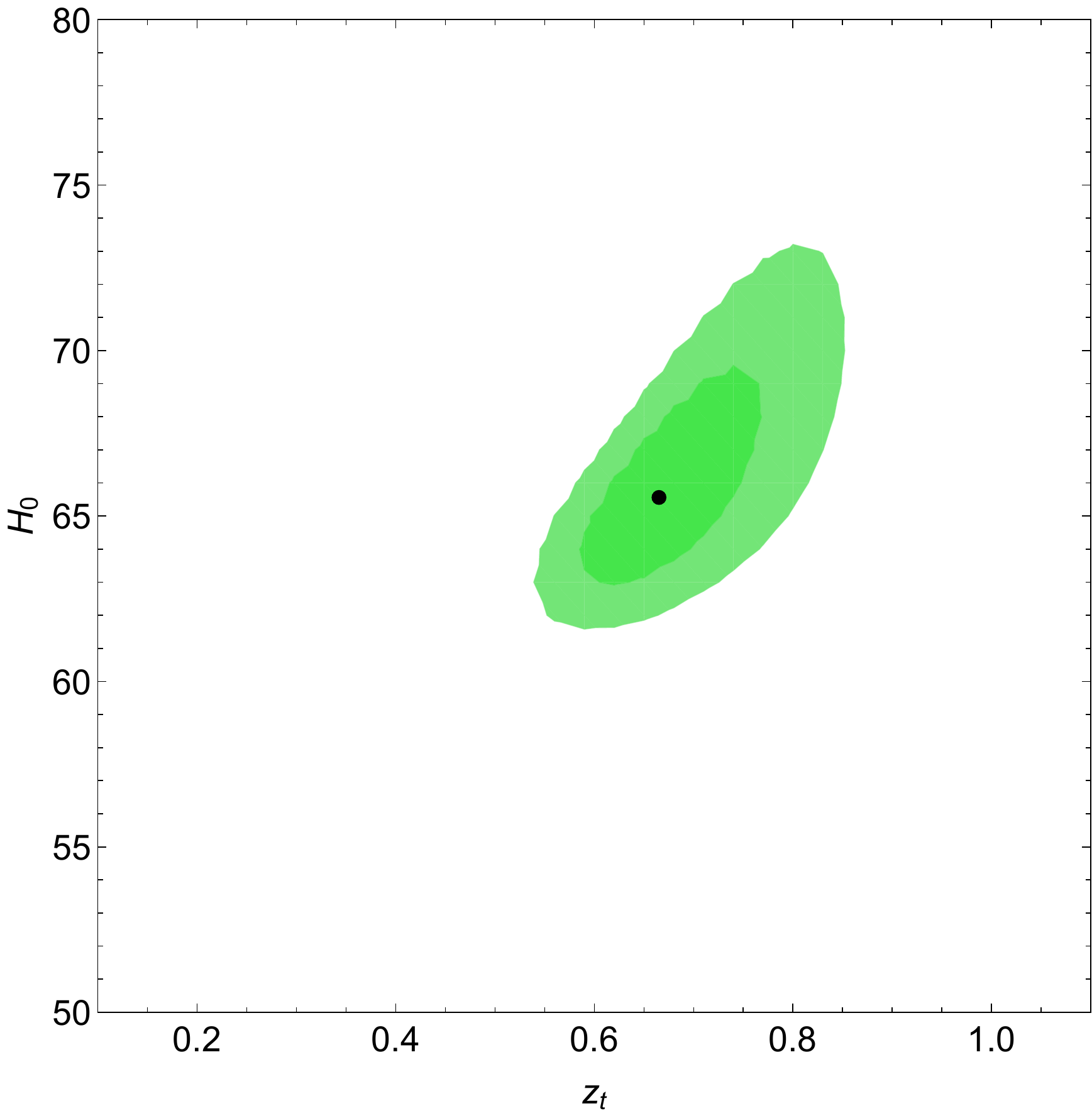}
		\includegraphics[scale=0.400]{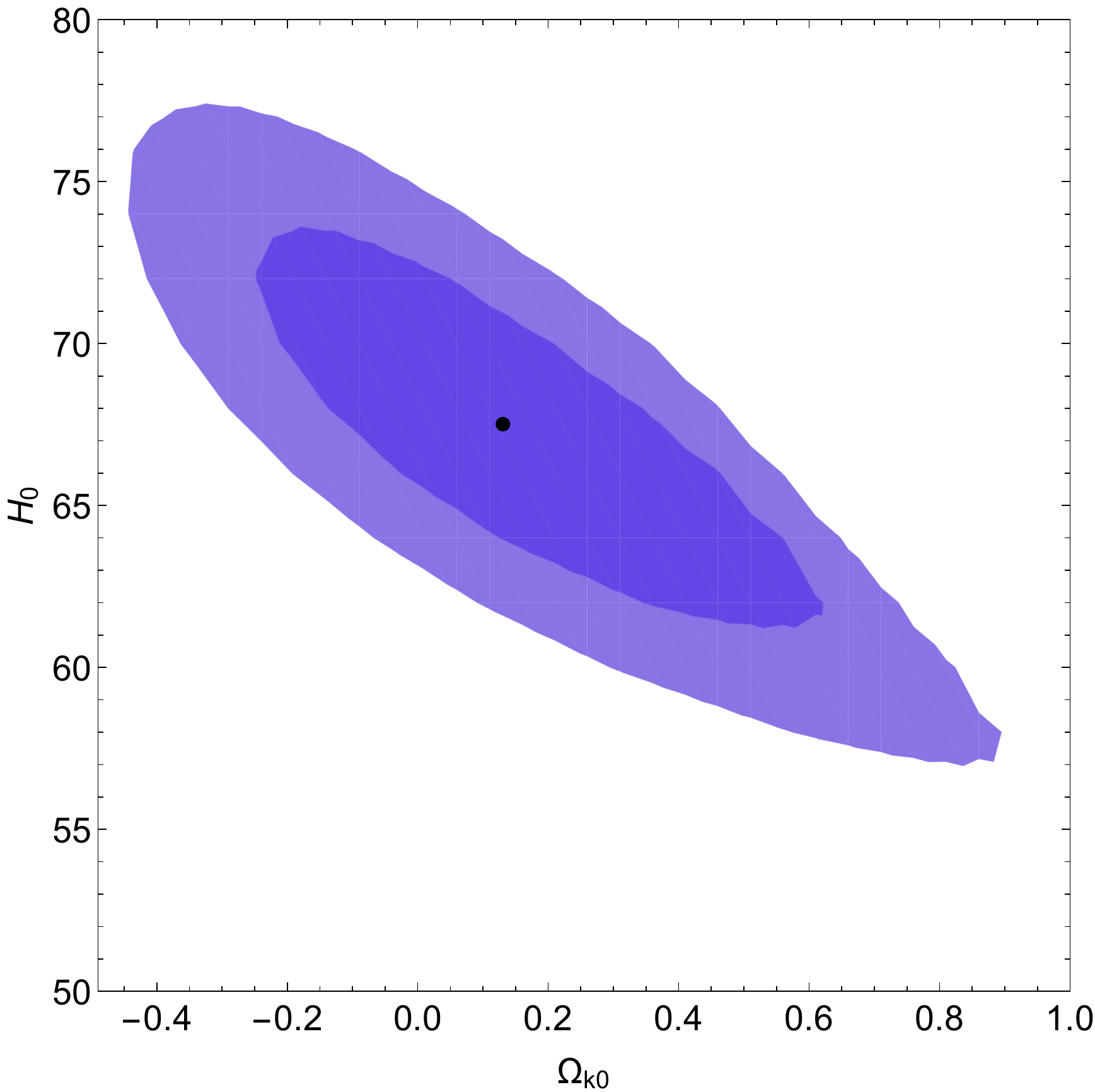}
 		\includegraphics[scale=0.400]{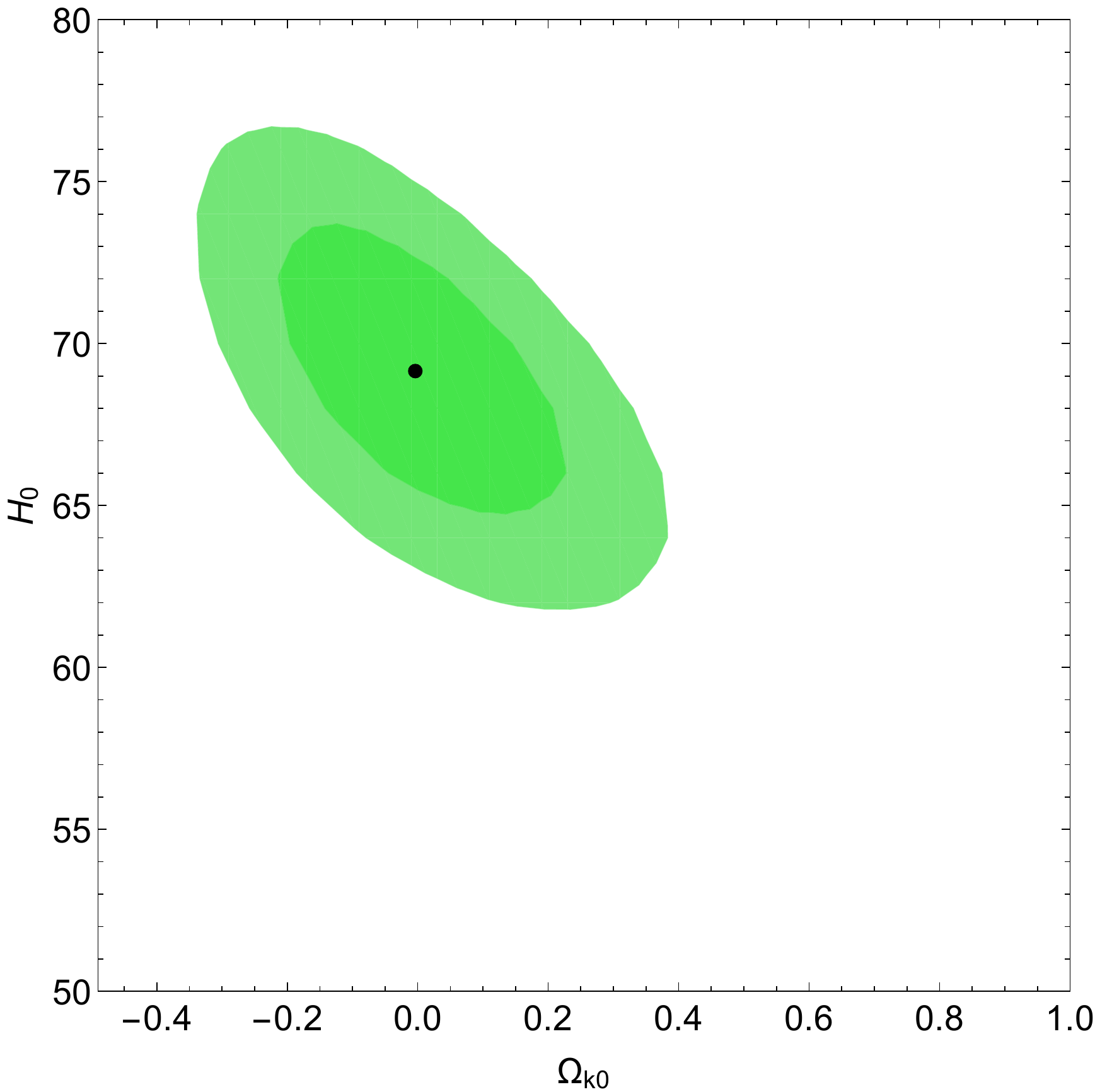}
 	\caption{In the figure on the left we show constraints on the Hubble constant using data from the Hubble parameter and on the left side we add data from Supernovas Ia. In all cases we marginalize the matter density parameter in the interval $ 0.07<\Omega_{m0}<0.6$.}
 	\label{fig01}
 \end{figure*}

\begin{figure*}[htbp] 
 	\centering
	\includegraphics[scale=0.300]{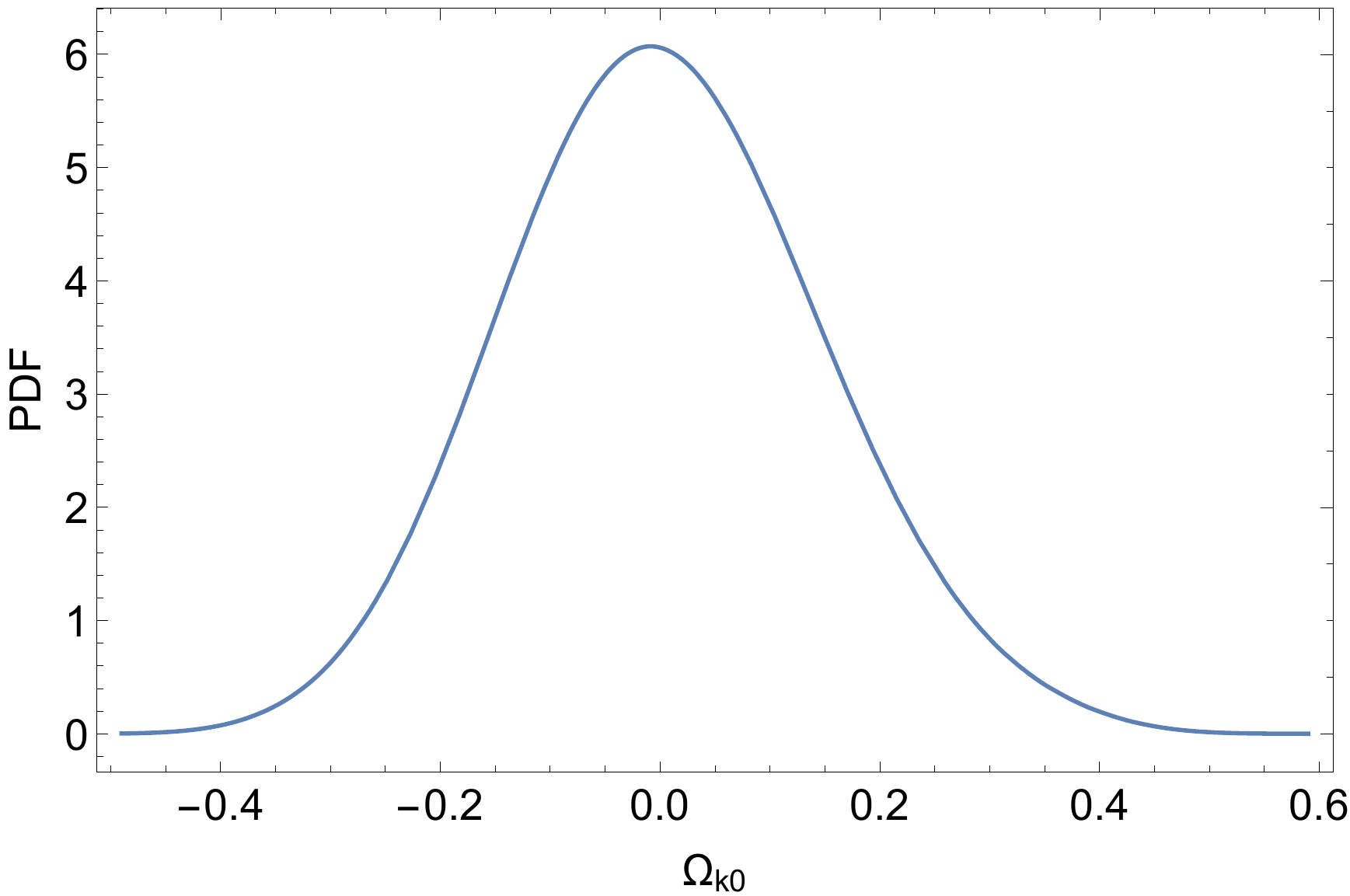}
 		\includegraphics[scale=0.300]{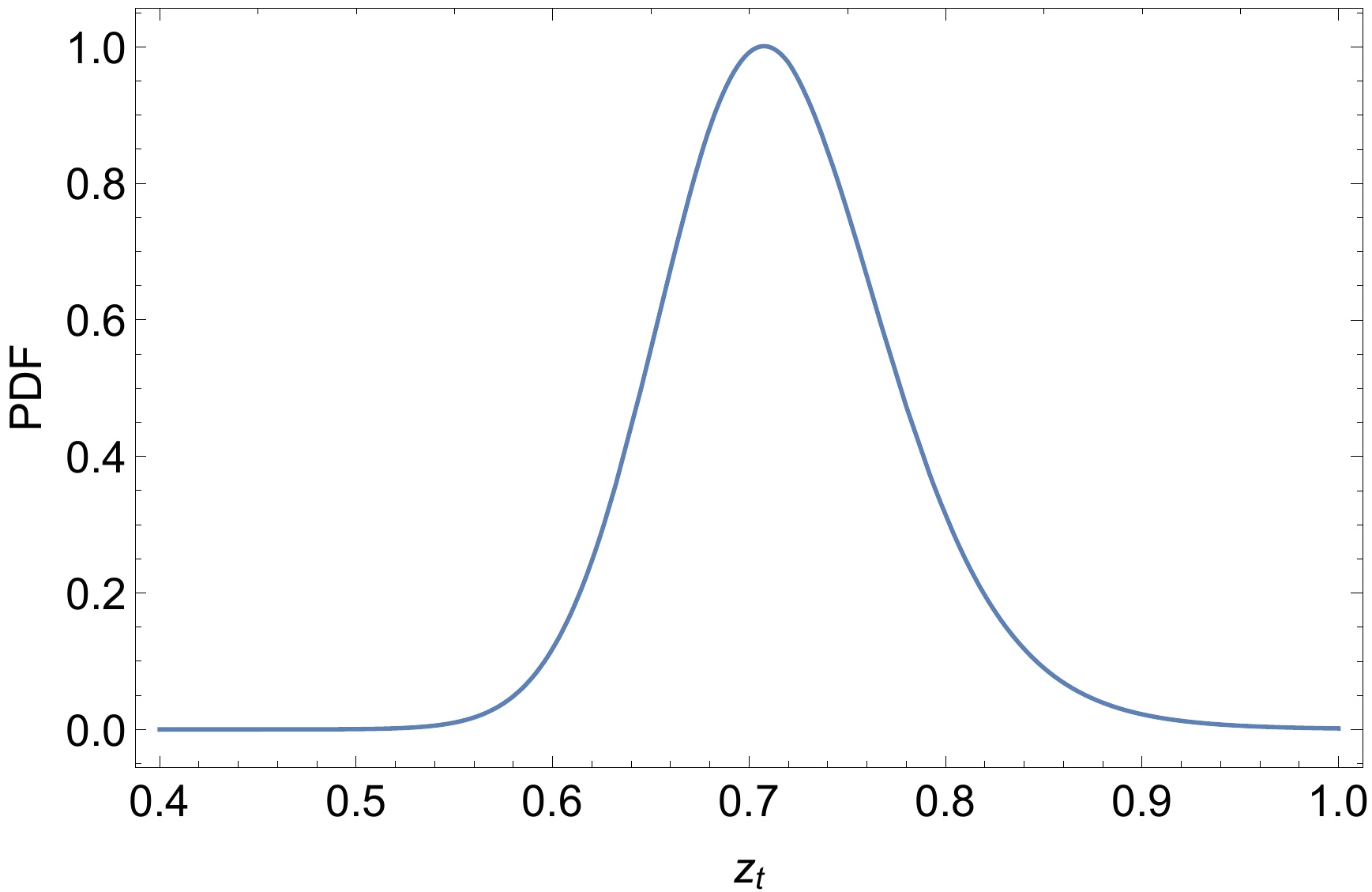}
		\includegraphics[scale=0.300]{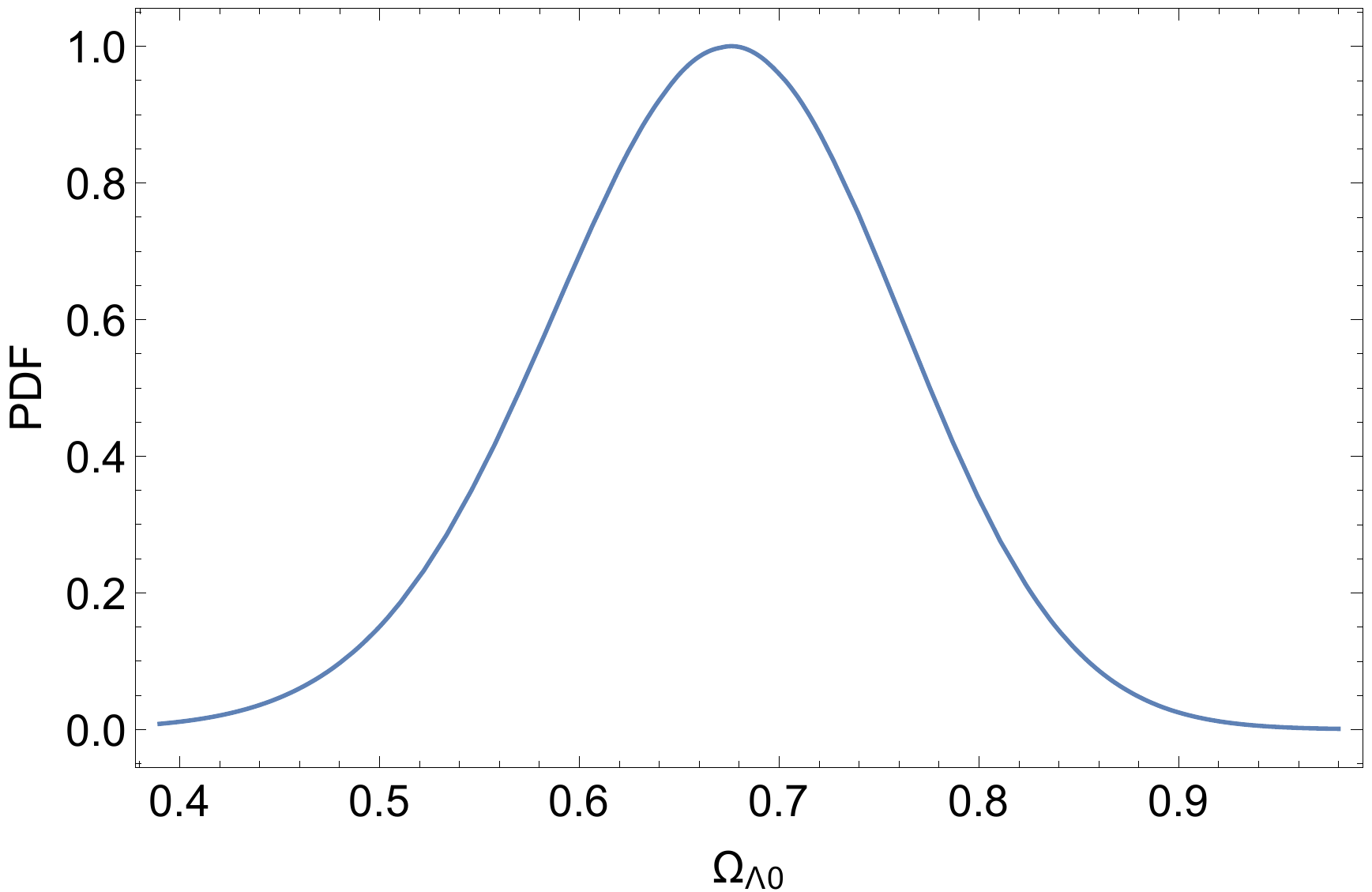}
 		\includegraphics[scale=0.300]{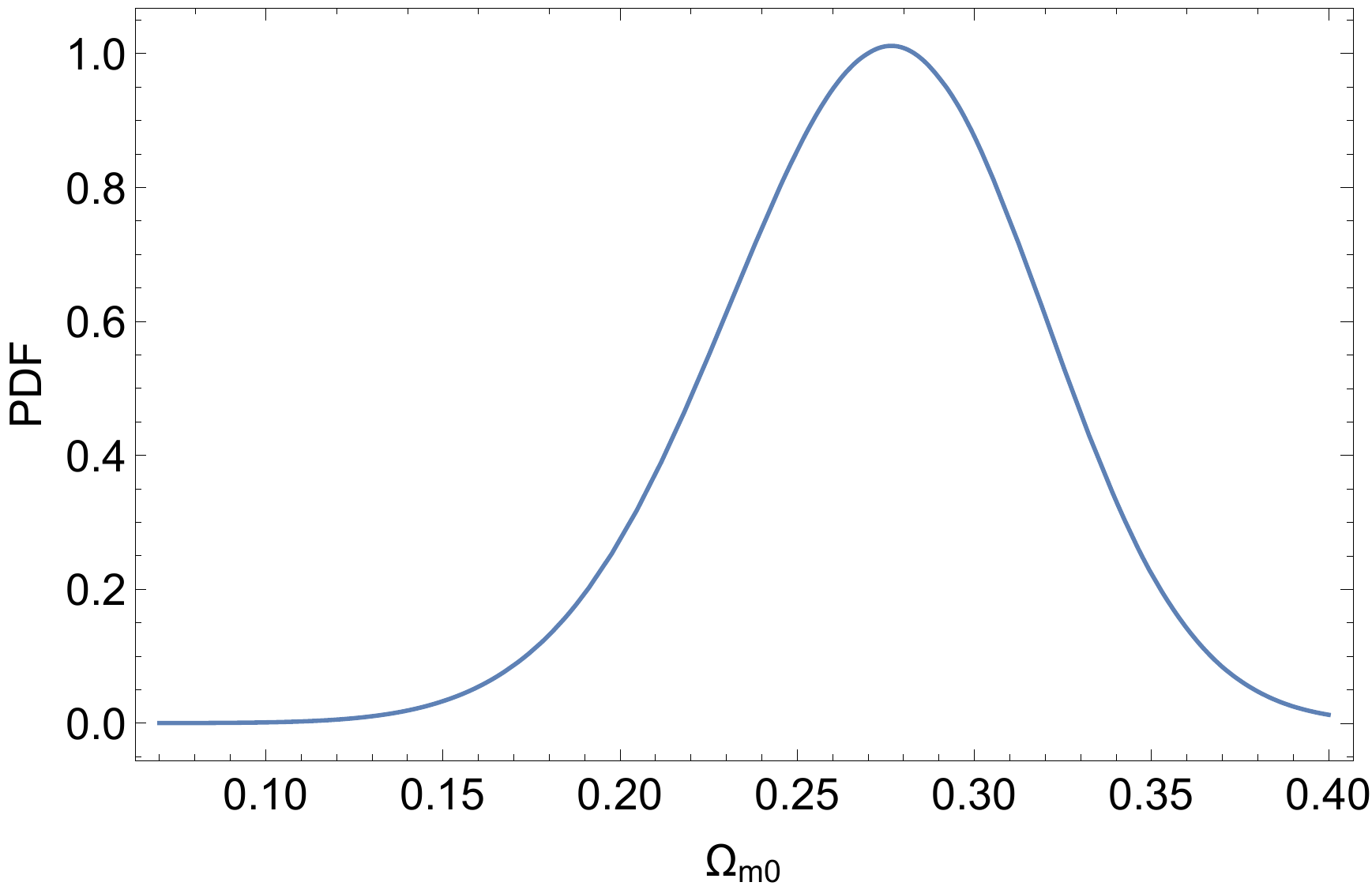}
			\includegraphics[scale=0.300]{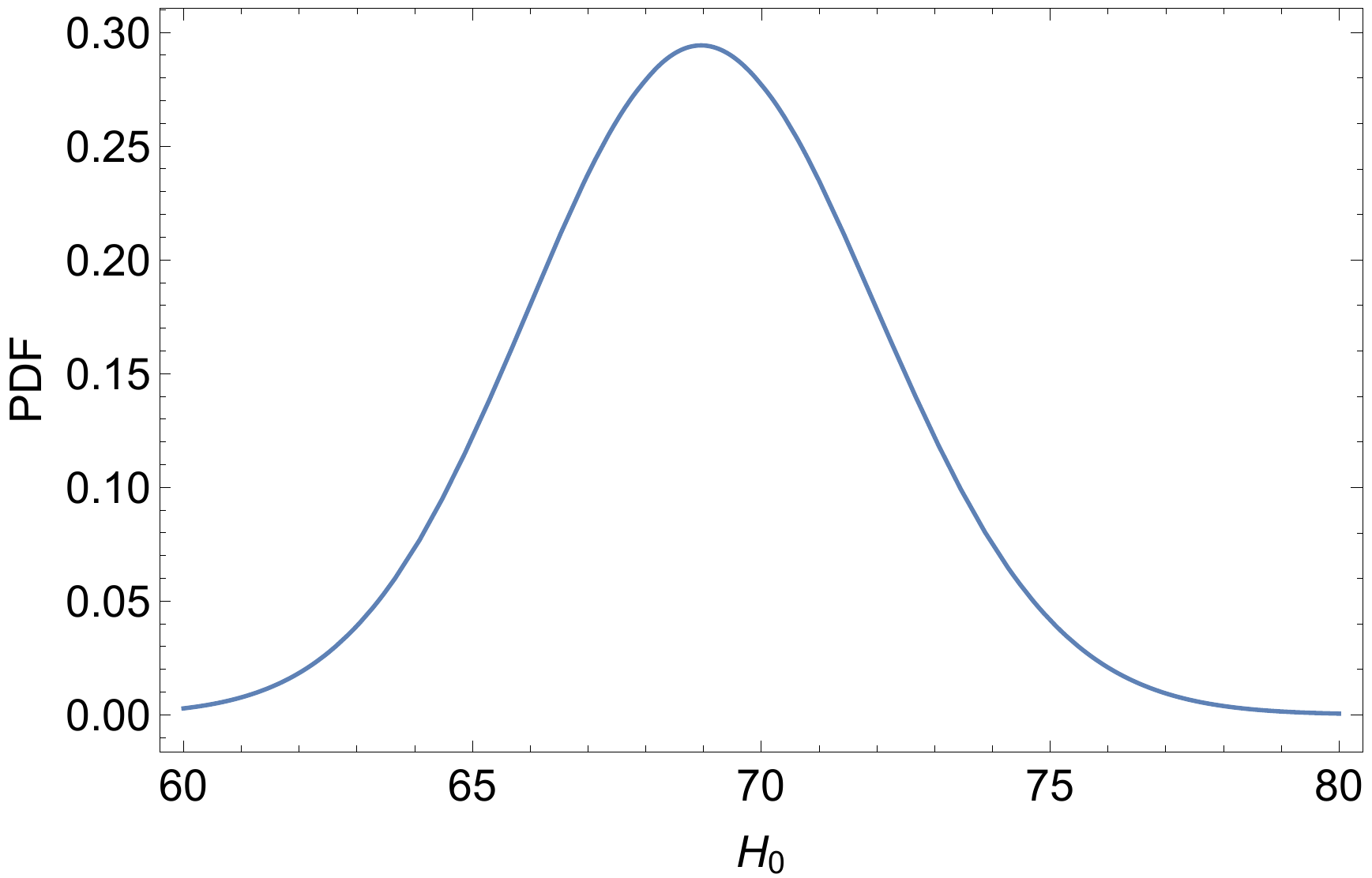}
 	\caption{In the figure we show the PDFs for all the observables involved in the $\Lambda CDM $ background model.}
 	\label{fig01}
 \end{figure*}

\begin{figure*}[htbp] 
 	\centering
	\includegraphics[scale=0.600]{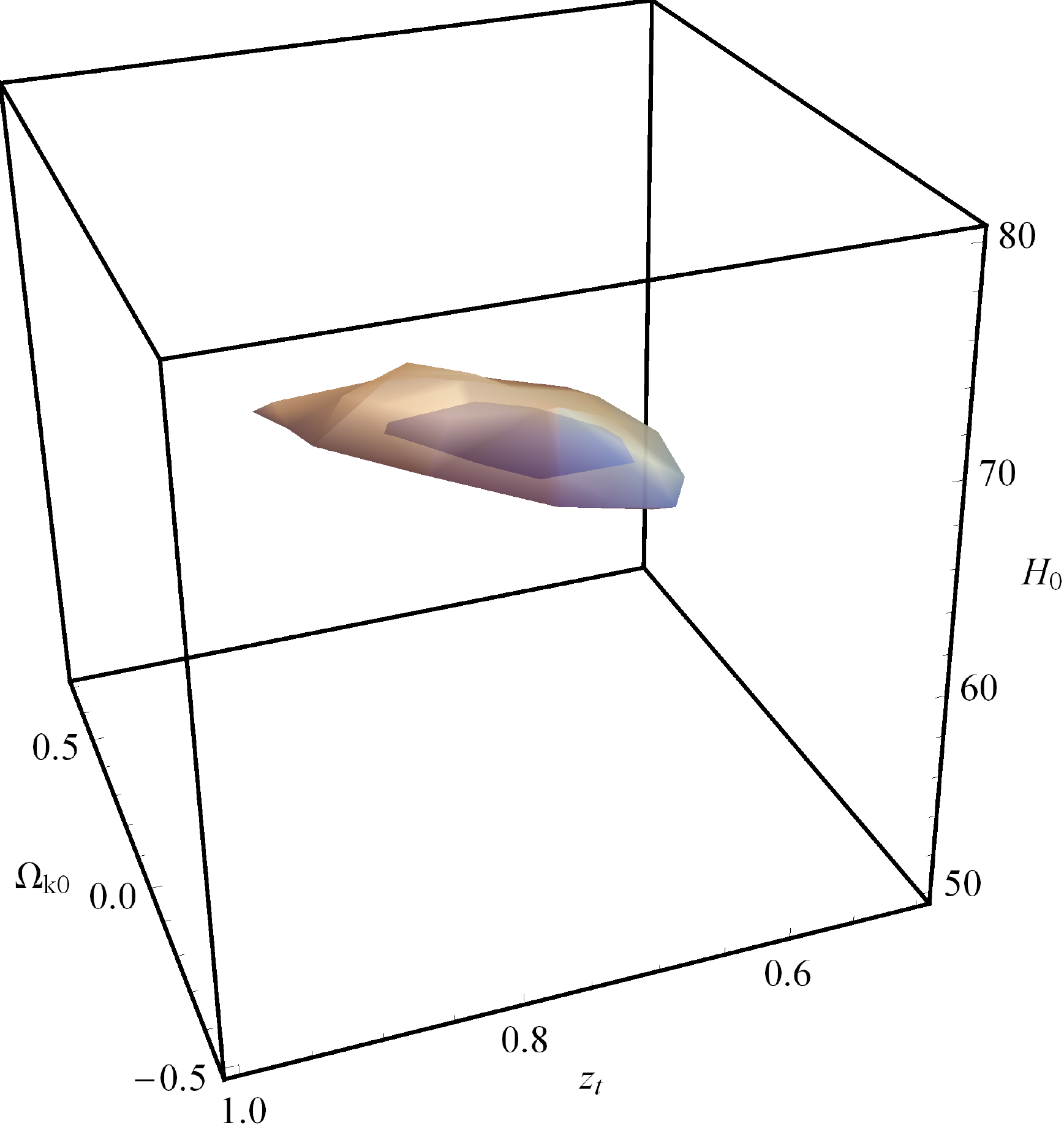}
 	\caption{ The $1\sigma$ - $2\sigma$ confidence contour of the parameter space $(z_{t},\Omega_{k0},H_{0})$ using 
	the joint analysis.}
 	\label{fig01}
 \end{figure*}

\begin{figure*}[htbp] 
 	\centering
	\includegraphics[scale=0.400]{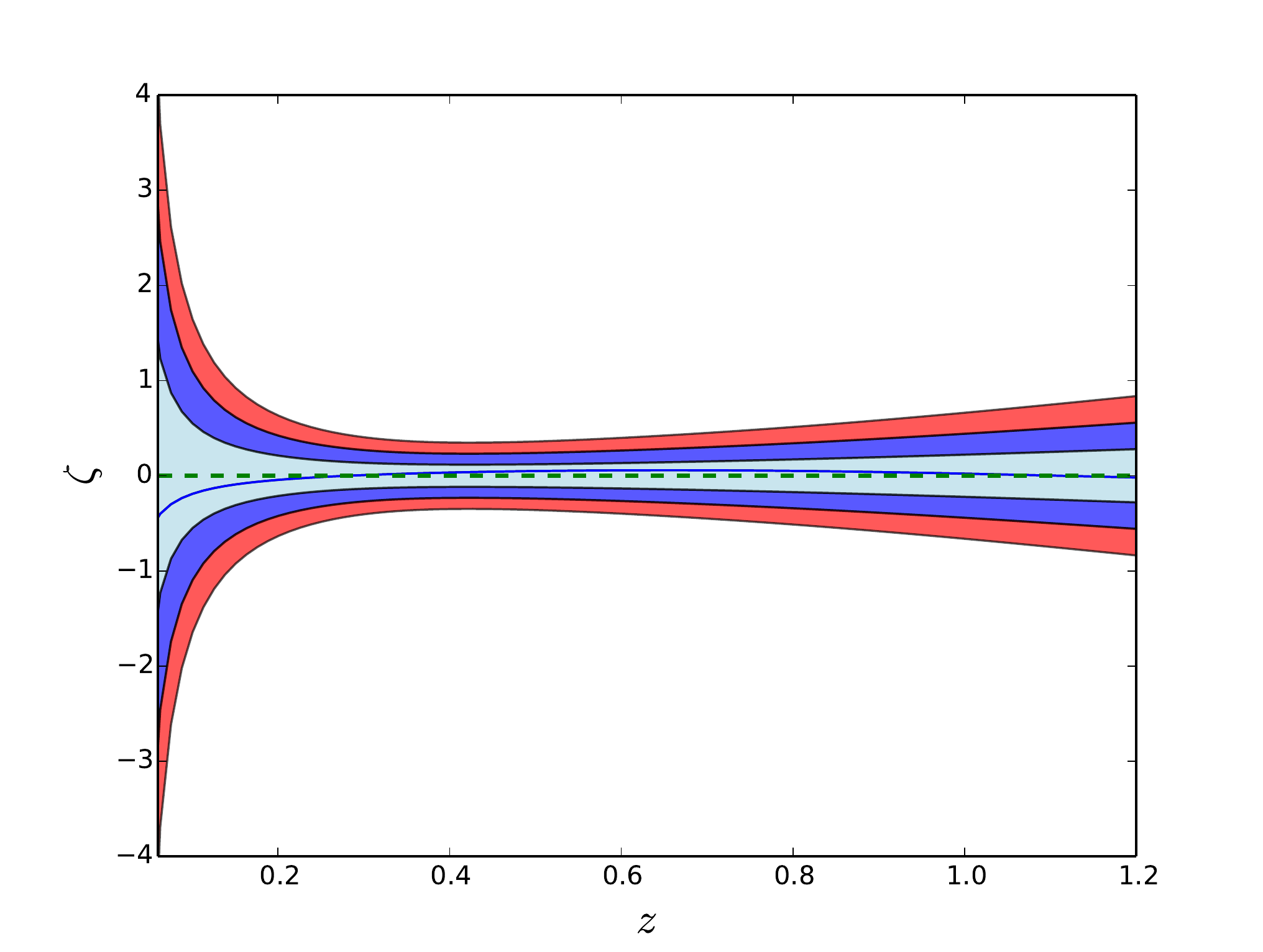}
	\includegraphics[scale=0.400]{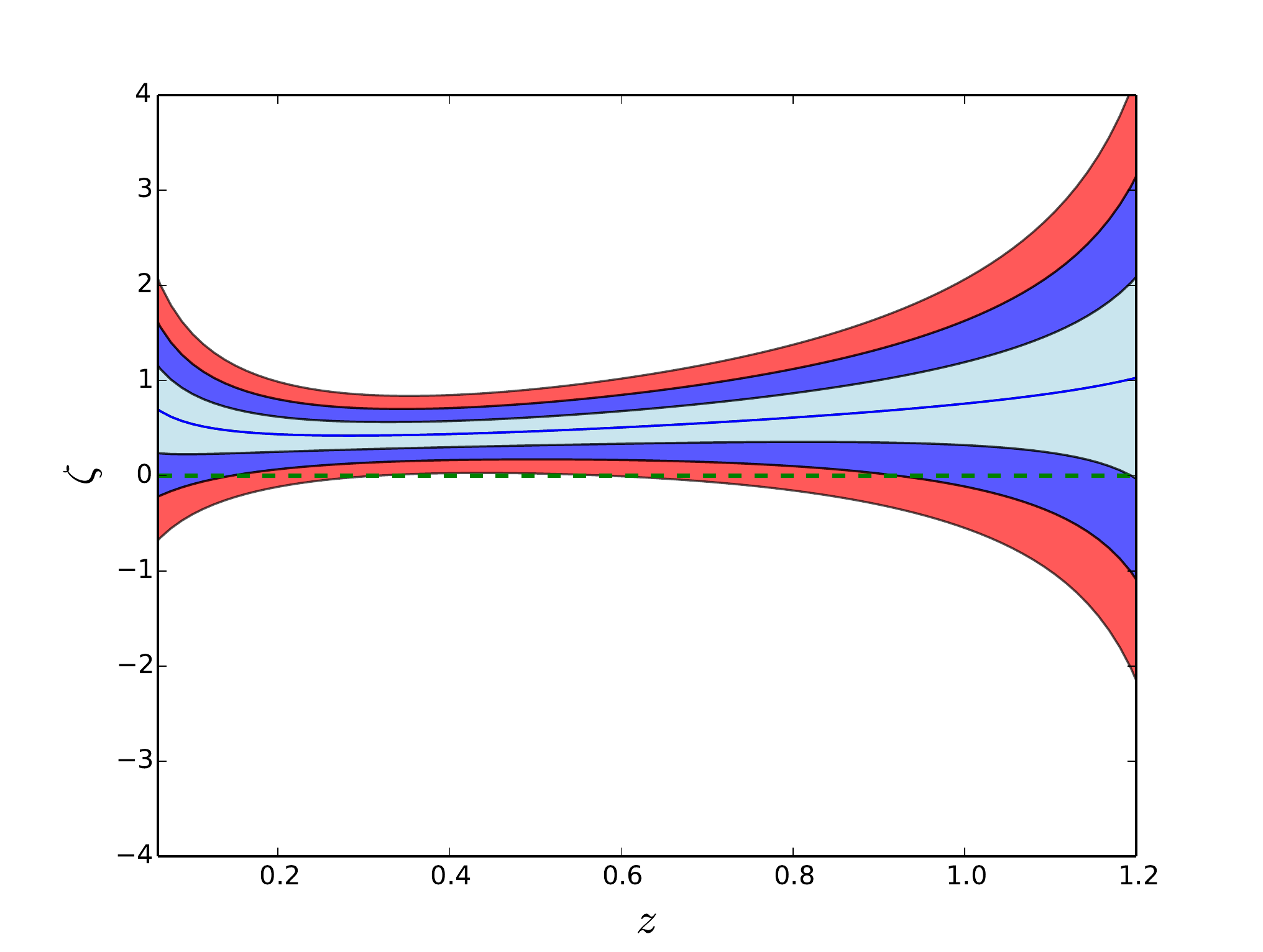}
	\includegraphics[scale=0.400]{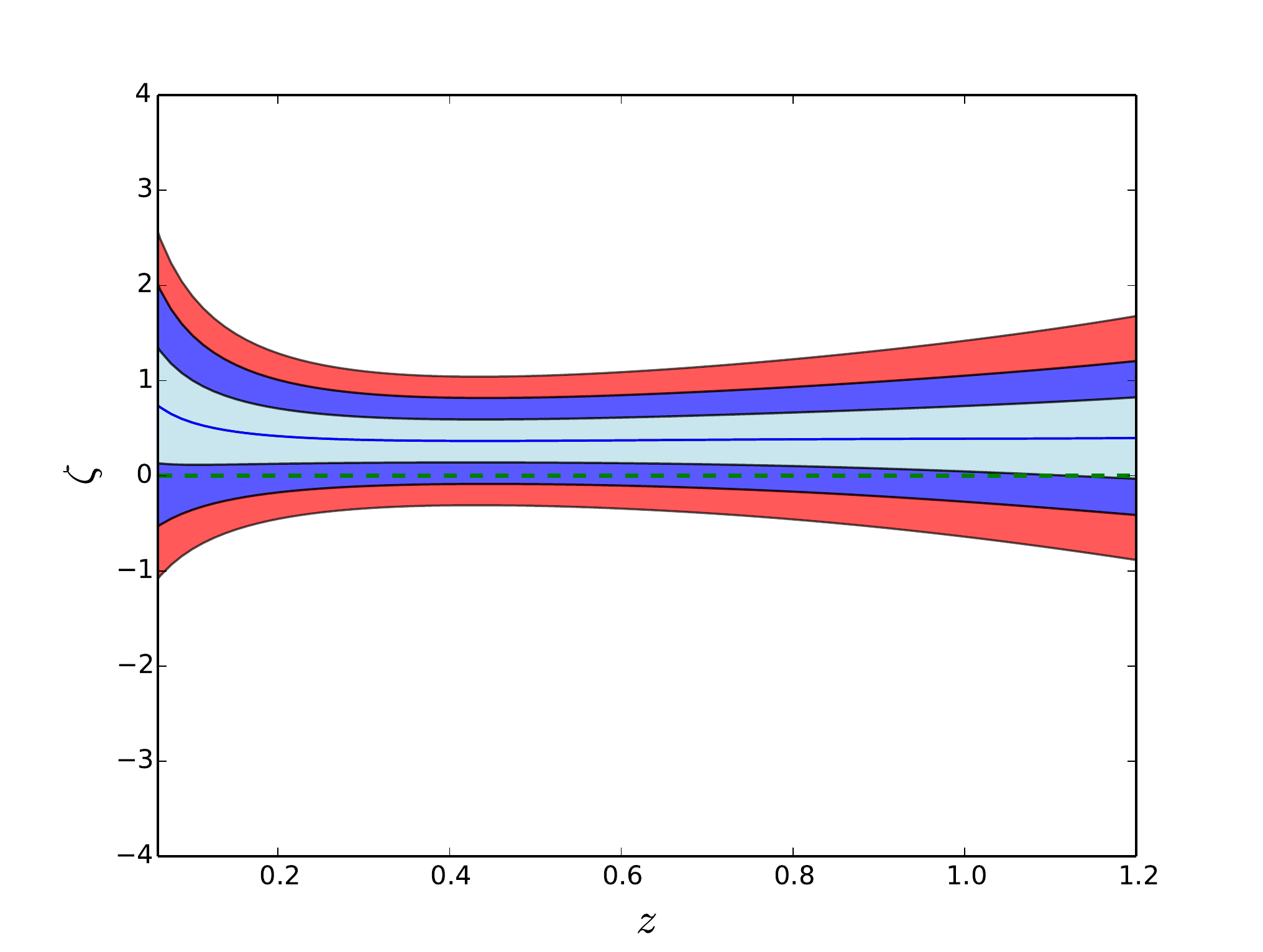}
 	\caption{The results for the reconstruction of the null test for the non-flat $\Lambda CDM$ using Hubble parameters measurements with $1\sigma$, $2\sigma$ and $3\sigma$ of C.L. In the figure above on the left we can see the reconstruction of the null test using GP and the 
best-fit of PLANCK/2018. In the figure above on the right we show the reconstruction using the $H_{0}74.03 \pm 1.42$ of RIESS et al./2018. In the figure below we maintain the value of the Hubble constant of RIESS et al. and use $\Omega_ {m0} = 0.28 \pm 0.01$. The dashed line represents the flat $\Lambda CDM$ model and the solid blue line represents the average value od the $\zeta$ test.}
 	\label{fig01}
 \end{figure*}

\section*{Acknowledgments}
A.M.V.T would like to thank the reading of the manuscript and suggestions made by J.C. Fabris. 
We also appreciate the computational facilities of the UFES to develop the work.

\end{document}